\documentclass{aa}  
\usepackage{natbib}
\usepackage{graphicx}
\usepackage{rotating,booktabs}
\usepackage[verbose]{placeins}
\usepackage{pdflscape}
\usepackage{tabularx}
\usepackage{adjustbox}
\usepackage{tikz}
\usepackage[english]{babel}
\usepackage{hyperref}
\hypersetup{colorlinks   = true, citecolor = blue,
            urlcolor=blue,
            linkcolor= blue}

\usepackage{txfonts}
\usepackage{longtable}

\newcommand{\lya}{\ion{Ly$\alpha$}{}}
\newcommand{\heii}{\ion{He}{II}}

\newcommand{\ha}{\ion{H$\alpha$}{}}
\newcommand{\hii}{\ion{H}{II}}
\newcommand{\heiii}{\ion{He}{III}}

\newcommand{\Laz}{\textcolor{blue}{Lazeyras et al., in prep.}}

\begin{document} 

   \title{Constraining the density distribution of the cold Circumgalactic Medium of quasars at $z>2$ through \lya, \heii\ and \ha\ emission}

   \titlerunning{MAGG}
   
   \author{A. Travascio
          \inst{1,2}
          \and
          S. Cantalupo
          \inst{2,1}
          \and
          T. Lazeyras
          \inst{2}
          \and
          Naveen A. Reddy
          \inst{3}
          \and
          C. C. Steidel
          \inst{4}
          \and
          Y. Liang
          \inst{5}
          \and
          G. Pezzulli
          \inst{6}
          \and
          M. Gaspari
          \inst{7}
          \and
           M. Fossati
          \inst{2,8}
          \and
          M. Fumagalli
          \inst{1,2}
          \and
          C. Feruglio
          \inst{1}
          \and
          F. Fiore
          \inst{1}
          \and
          H. Mao
          \inst{9}
          \and
          M. Galbiati
          \inst{1}
          \and
          W. Wang
          \inst{2}
          \and
          N. Ledos
          \inst{2}
          \and
          B. Mobasher
          \inst{3}
          \and
          A. Pensabene
          \inst{10}
          \and
          G. Quadri
          \inst{2}
          }

\institute{INAF – Osservatorio Astronomico di Trieste, Via G. B. Tiepolo 11, 34143 Trieste, Italy \\ e-mail: \texttt{andrea.travascio@inaf.it}  \and
    Dipartimento di Fisica "G. Occhialini", Università degli Studi di Milano-Bicocca, Piazza della Scienza 3, I-20126, Milano, Italy \and 
    Department of Physics and Astronomy, University of California, Riverside, 900 University Avenue, Riverside, CA 92521, USA \and 
    Cahill Center for Astrophysics, California Institute of Technology, Pasadena, CA 91125, USA \and
    Institute for Cosmic Ray Research, The University of Tokyo, 5-1-5 Kashiwanoha, Kashiwa, Chiba 277-8582, Japan \and
    Kapteyn Astronomical Institute, University of Groningen, Landleven 12, 9747 AD Groningen, The Netherlands \and
    Department of Physics, Informatics and Mathematics, University of Modena and Reggio Emilia, 41125 Modena, Italy \and
    INAF - Osservatorio Astronomico di Brera, via Brera 28, 20121 Milano, Italy \and
    Purple Mountain Observatory, Chinese Academy of Sciences, 10 Yuanhua Road, Nanjing 210023, People's Republic of China 2 \and 
     DTU Space, Technical University of Denmark, DK-2800 Kgs. Lyngby, Denmark
}

\abstract
{Extended \lya\ emission is routinely detected around high-redshift quasars, revealing large reservoirs of cool circumgalactic medium (CGM). However, its resonant nature complicates the interpretation of the gas density, ionization state, and kinematics, making non-resonant tracers such as \heii\ and \ha\ essential. 
Extending previous analytical models, we perform a suite of \texttt{CLOUDY} photo-ionization simulations to predict the \heii/\ha\ ratio in quasar-illuminated gas, considering a large range of physical parameters and including log-normal density distributions. These models demonstrate that the \heii/\ha\ ratio can effectively constrain the amplitude of density fluctuations in the emitting gas, commonly quantified by the clumping factor, $C$.
We then compare these predictions with observations of extended \lya, \heii, and \ha emission around quasars at $z\simeq2.1-4.5$. At $z\approx2.2$, we observed a sample of 13 quasars with Keck/KCWI and obtained targeted Keck/MOSFIRE follow-up observations for the systems showing the most prominent extended \lya\ and \heii\ emission. This provides the first statistical sample of quasar CGM systems with combined \lya, \heii, and \ha\ measurements.  
Within matched apertures extending to approximately $35~\rm pkpc$, stacked spectra from these observations yield $\lya/\ha \approx 9.1_{-3.2}^{+5.3}$, $\heii/\ha =0.54^{+0.32}_{-0.21}$, and $\heii/\lya =0.06_{-0.02}^{+0.03}$. 
In single-density models, such a low \heii/\ha\ ratio would require extremely high CGM densities ($n>10$ cm$^{-3}$) or implausibly soft AGN spectra. By contrast, broad log-normal density distributions can reproduce the observed ratio for lower median densities, but require a substantial emissivity-weighted high-density tail; in the adopted lognormal parameterization, this corresponds formally to a clumping factor of order $\approx 10^3-10^4$. 
We further analyze 26 VLT/MUSE observations of quasars at $3.1<z<4.5$, for which \ha\ observations are not possible from the ground, finding 14 systems with co-spatial extended \lya\ and \heii\ emission. The detected, emission-weighted \heii/\lya\ component shows no evidence for strong evolution between $z\sim 2.2$ and $z\sim 3.7$, although this does not imply an invariant underlying density PDF.
By providing new observational constraints, our results can be used together with theoretical models and hydrodynamical simulations in order to advance our knowledge on the physical processes shaping the CGM of quasars at high redshifts. 
}


   \maketitle

\section{Introduction}

The standard cosmological model of large-scale structure formation predicts that both the dark matter and the gas are distributed within interconnecting filaments, sheets, and over-density regions (i.e. nodes), which compose the ''Cosmic Web'' structure \citep{Bond96}. 

Observational studies of the density of diffuse gas in the Cosmic Web, pioneered by \citet{Sargent80}, have identified two distinct environments: the CircumGalactic Medium (CGM) and the InterGalactic Medium (IGM). The IGM consists of gas with low density distributed along filaments that connect regions of higher density. These filaments form the backbone of the large-scale structure of the universe, connecting galaxy clusters and galaxy groups. The CGM refers to the diffuse gas residing within a few virial radii of a galaxy's dark matter halo.

These baryonic structures can be studied through absorption features in continuum emission of background quasars \citep[e.g.][]{Adelberger03, Fumagalli13,Prochaska13,Prochaska18}, tracing different density regions of the cool diffuse gas and the associated metals. 
On the other hand, well-known analytical predictions, confirmed by cosmological simulations including radiative transfer effects, suggest that self-shielded CGM, exposed to the metagalactic UV background, should emit \lya\ photons as most prominent transition with $\rm{SB}_{\lya} \approx 10^{-20} \rm ~erg~s^{-1}~cm^{-2}~arcsec^{-2}$, i.e., $\sim$10 times fainter than the sensitivity limits of current instruments \citep[e.g.][]{Gould96,Cantalupo05,Kollmeier10,Cantalupo12}. This value is determined primarily by the UV background intensity. Therefore, in proximity to bright UV sources, such as quasars, the emission of \lya\ photons from recombination is expected to increase, leading to easily detectable values such as $\rm{SB}_{\lya} \geq 10^{-18}~\rm erg~s^{-1}~cm^{-2}~arcsec^{-2}$. 

In accordance with these expectations, in recent years, there have been several detections of the \lya\ transition of the diffuse hydrogen distributed at distances $> 50~\rm pkpc$ around bright radio-quiet quasars at $z>$2. These detections, known as \lya\ nebulae, provide insights into the densest regions of the Cosmic Web structure \citep[i.e., CGM and partially IGM; see][]{Cantalupo14,Hennawi15,Borisova16,Battaia19,Fossati21,Tornotti25}.\\

Characterizing this diffuse gas, especially in the outskirts of galaxies, is essential for understanding galaxy evolution, because this material plays a key role in regulating gas accretion and feedback processes.
In this context, absorption and emission observations offer complementary constraints on the density structure of the CGM, allowing us to investigate whether this gas is smoothly distributed or fragmented into dense unresolved clumps, commonly quantified by the clumping factor, $C= \langle n^2 \rangle / \langle n \rangle ^2$, describing the density fluctuations of the gas. Absorption studies indicate that the cool CGM is multiphase and highly inhomogeneous, with dense small-scale structures embedded in more diffuse gas \citep{Stern16}. 
In parallel, extended \lya\ emission around galaxies and quasars shows that large reservoirs of cold gas can be detected on CGM scales, although reproducing the observed surface brightness (SB) often requires this gas to have high densities ($n\sim 0.1-1~\rm cm^{-3}$) or large density fluctuations \citep[C$>$100;][]{Cantalupo14,Wisotzki16}.

While \lya\ is the brightest and most commonly detected tracer of the cool CGM, its resonant nature and multiple production mechanisms introduce strong degeneracies among the gas density, ionization state, geometry, and kinematics. Indeed, \lya\ photons can be produced locally by atomic recombinations of ionized gas, collisional excitation (sometimes called "cooling radiation"), and photon-pumping from bright sources (sometimes called "scattering"), see \citealt{Cantalupo17} for a review. At the same time, once produced, \lya\ photons can be redistributed over large spatial scales and different frequencies by resonant scattering. Thus, non-resonant recombination lines, such as \ha, and high-ionization tracers, such as \heii, are needed to provide vital constraints on the ionization state, density, and geometry of the emitting gas. 

In particular, by using analytical photo-ionization models, \cite{Cantalupo19} showed that the ratio of two recombination lines from hydrogen and helium could be used to infer the broadness of the emitting gas density distribution and applied this methodology to \lya\ and \heii\ integral-field spectroscopic observations of extended gas in the Slug Nebula at $z\sim2.3$. This model assumed that \lya\ is mostly produced by recombinations, as suggested by previous MOSFIRE long-slit observations of \ha emission for this system \citep{Leibler18}. The comparison of \lya, \heii\ and \ha emission in this complex system suggested the presence of multiple structures oriented along the line of sight and that only the most distant region (dubbed 'bright tail') was covered by the MOSFIRE \ha\ longslit observations at sufficient depth to ensure a significant detection (despite the presence of a sky emission line at the same wavelength). The inferred distance from the ionizing source of such a region would put it well beyond the CGM of the bright quasar, that is responsible for the gas 'illumination'. A subsequent experiment performed by \cite{Langen23}, who observed the bright quasar J0010+06 at $z\sim2.26$ using KCWI and MOSFIRE longslit observations, detected both \lya and \ha emission from the quasar CGM, providing among the first pieces of evidence that \lya emission on quasar CGM scales is consistent with a recombination radiation origin. 

While \ha\ observations are possible only within small redshift windows, e.g., at $z\simeq2.3$, from the ground or require space-based IR observations, \heii\ is a UV line that can be detected with MUSE and KCWI for a large redshift range. 
Despite that, observations of \heii\ nebulae are sparse and are typically associated with individual extreme environments hosting enormous and highly luminous \lya\ nebulae \citep{Cai17,Marino19,Cantalupo19,Chaves19,Herenz20,Andrade22}. In other cases, the \heii\ emission is present at a very faint level, resulting in upper limits \citep{Borisova16,Battaia19}.

The goal of this work is to improve over previous studies both on the modelling and on the observational sides by providing: i) a refined photo-ionization model to predict the expected \heii\ to \ha\ ratios as a function of a large range of parameters using CLOUDY simulations \citep{Ferland98}; ii) a specifically selected sample of quasar CGM observations at $z\sim2.3$ with KCWI and MOSFIRE providing, at the same time, \lya, \heii\ and \ha observations; iii) a large statistical sample of individually detected \heii\ nebulae associated with bright quasar \lya nebulae at $z>3$ detected with MUSE. By comparing the results of our photo-ionization models with this first set of observations, our work provides, for the first time, robust constraints on both the density distribution of the CGM of quasars and on the emission mechanism of their extended \lya emission. 

In particular, the paper is structured as follows. In Section~\ref{sec:cloudy} we present a suite of \texttt{CLOUDY} photo-ionization models and derive the expected \heii/\ha\ line ratios for different gas density distributions and ionizing spectra. In Section~\ref{sec:z22}, we describe the Keck/KCWI and Keck/MOSFIRE observations targeting $z \simeq 2.2$ quasars, their data reduction and analysis aimed at extracting faint line-emitting CGM, and measure the \lya, \heii, and \ha\ line ratios. In Section~\ref{sec:z34}, we extend the analysis to a sample of VLT/MUSE observations of $z \simeq 3.1-4.5$ quasars and compare their \heii/\lya\ line ratios with those around $z \simeq 2.2$ quasars in the KCWI sample. Finally, in Section~\ref{sec:obsmodels}, we compare the \heii/\ha\ ratios observationally constrained with the \texttt{CLOUDY} predictions to infer the median density distribution and the typical clumping factor of the quasars CGM, and our summary and conclusions are reported in Section~\ref{sec:sumcon}.

Throughout this work we assume the cosmological parameters $H_0 = 67.4 \rm ~km ~s^{-1}~Mpc^{-1}$ and $\Omega_m = 0.315$ \citep{Plank18}. Unless otherwise noted, all uncertainties shown in the plots correspond to the 1$\sigma$ (68$\%$) confidence level.

\section{Analytical Expectations from \texttt{CLOUDY} modeling}\label{sec:cloudy}

In this section, we describe the main ingredients of our photo-ionization model aimed at providing detailed constraints on the expected \heii\ over \ha\ ratios for different gas density distributions and as a function of several other parameters related to both the emitting cloud properties and the quasar ionizing radiation spectrum. In particular, our model is an extension of the analytical model developed in \cite{Cantalupo19} in which we allow for temperature variations, the presence of metals, and the helium self-shielding effect, which require more sophisticated numerical modelling, such as CLOUDY simulations. 
At the same time, we focus on the \heii/\ha\ line ratio rather than on the \lya\ transition, whose resonant nature and sensitivity to radiative transfer effects make it considerably more difficult to model reliably within standard \texttt{CLOUDY} photo-ionization calculations. Assuming case B recombination and a temperature of $2\times10^4$ K, our results could be converted into a \heii/\lya\ ratio by dividing the \heii\ to \ha\ ratio by a factor of about 8.3 \citep[see][for details]{Langen23}. However, this conversion is only valid in the absence of radiative transfer effects.

Our models assume that the \heii\ and hydrogen emission lines are produced by the same clouds and are therefore co-spatial. This is a good assumption as demonstrated by the comparison of the \heii\ and \lya\ SB profiles of observed nebulae at z$>2$ (Appendix~\ref{app:spec}). Furthermore, we assume that the \ha\ and \heii\ emitting CGM is composed of warm (T$\sim10^4~\rm K$) gas clouds, with a given density and size, illuminated by the quasar radiation.
To reflect the typical scales probed by recent CGM surveys, we adopt physical distances from the quasar ($D_c$) of 10 and 50 kpc, corresponding to the inner and outer limits where non-resonant lines are most commonly detected, both in previous studies and in our data (Figures~\ref{appfig:slits}, \ref{figure:SBmaps}). We adopt fiducial cloud sizes of $l_{\rm c}=70$ and 200 pc, corresponding to plausible combinations of volume filling factor ($f_V$), covering factor ($f_C$), and halo mass (Eq.~14 of \citealt{deBeer23}). We also explored $l_{\rm c}=1$ pc to 5 kpc, formally consistent with \citet{deBeer23}, but excluded these extremes because smaller clouds systematically increase the predicted \heii/\ha\ ratios, while kpc-scale clouds imply proportionally larger $f_V$, exceeding the small filling factors inferred for the multiphase CGM \citep[$f_V<10^{-2}$, with typical values $\langle f_V \simeq 10^{-3}$][]{Prochaska09,Pezzulli19,Liang20}.

A crucial feature of our models is that we allow the gas temperature to vary freely and self-consistently throughout the cloud (see Appendix~\ref{sec:temp} for details about $T$ versus the hydrogen density). This avoids the potential effect of the simplifying assumption of a fixed temperature, which could alter the predicted \heii/\ha\ ratio due to the different temperature dependencies of the helium and hydrogen recombination coefficients. 
We perform our CLOUDY simulation over a grid of hydrogen densities ($n_{\rm H}$) ranging from $10^{-5}$ to $10^5~\rm cm^{-3}$ with a step of 0.125 dex, covering the transition from the diffuse IGM to dense, optically thick clouds. We fixed the gas metallicity to $\log(Z/Z_{\odot})=~$-1 \citep{Fumagalli16b}, as varying it by one dex typically changes the predicted \heii/\ha\ ratio by $\lesssim10\%$, much less than variations driven by the cloud distance from the quasar. 

The AGN spectral energy distribution (SED) is modeled by two power laws in the energy range $8 \rm ~eV \leq E \leq 10 ~keV$. One power law between 2-10 keV assumes a slope of $f_{\nu} \propto \nu ^{-1}$. We note that this choice does not significantly affect our results, since the \heii\ and hydrogen ionization state is primarily sensitive to photons in the $\sim 1-4$~Ryd range.
To test the sensitivity of the \heii/\ha\ ratio to the extreme ultraviolet (EUV) radiation, we manually vary the slope $\alpha_{EUV}$ (defined at $\lambda < 912$~\AA). This allows us to bridge the UV and X-ray regimes with different degrees of hardness. We consider two representative EUV slopes: $\alpha_{EUV} = -1.7$ and $-2.2$. The former is commonly adopted and motivated by stacked quasar spectra \citep{Lusso15}, and we will consider this as our fiducial value. Although a value of $\alpha_{EUV} = -2.2$ is softer than currently allowed by, e.g., Helium reionization constraints ($-1.6<\alpha_{EUV}<-2.0$; \citealt{Khaire17}), we will nonetheless use this value as an extreme case for testing the impact of reduced high-energy photon production on the predicted line ratios. The spectrum is normalized at $0.6284$~Ryd based on the typical range of absolute magnitudes $M_{1450} \approx -29$ and $-27$ \citep[e.g.,][]{Fossati21}, which corresponds to average luminosities $\log(\nu L_{\nu} / \text{erg\,s}^{-1}) = 47.5$ and $46.6$ at $1450~\AA$, respectively. Finally, we include the cosmic ultraviolet background (UVB) as parameterized by \cite{Haardt12}, although at the distances considered in this study, its value is always subdominant with respect to the quasar radiation. Finally, our \texttt{CLOUDY} simulations compute the total emergent luminosities of \heii\ and \ha\ by integrating the local emissivity over the entire volume of the photoionized gas.

\subsection{Single Density Scenario}

\begin{figure}[t]
   \begin{center}
   \includegraphics[height=0.45\textheight,angle=0]{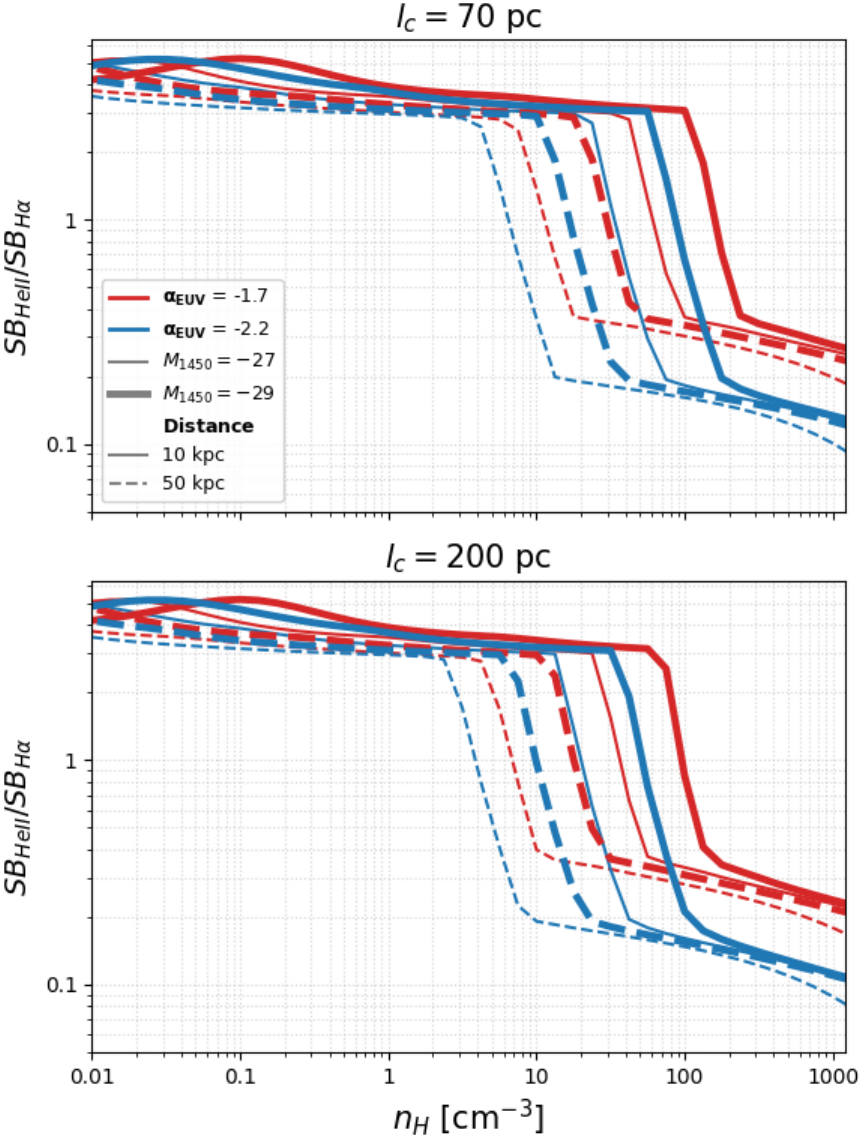} 
   \caption{\heii/\ha\ emission-line ratio as a function of gas density $n_{\rm H}$, computed from \texttt{CLOUDY} photo-ionization models assuming constant-density clouds. The models are shown for different EUV spectral slopes $\alpha_{\rm EUV}$, absolute magnitudes $M_{1450}$, cloud distances $D_{\rm c}$, and cloud sizes $l_c$ (top and bottom panels).}
      \label{figure:constDen}
   \end{center}
\end{figure}
\setlength{\textfloatsep}{-2pt}

Here, as an illustrative case, we explore the simplest scenario in which a single, uniform density characterizes the gas clouds. Figure~\ref{figure:constDen} shows the predicted $\rm SB_{\heii}/SB_{\ha}$ (equivalent to $\rm L_{\heii}/L_{\ha}$, since both are computed over the same emitting area and from the same emitting source) as a function of the hydrogen density ($n_{\rm H}$), for cloud sizes of $l_c = 70~\rm pc$ (top panel) and $l_c = 200~\rm pc$ (bottom panel). 
Solid and dashed lines correspond to different distances from the ionizing source ($D=10$ and $50$~kpc, respectively), while colors indicate the EUV spectral slope, $\alpha_{\rm EUV} = -1.7$ and $-2.2$, with the first one representing our fiducial value. The line thickness indicates two different AGN luminosities at rest-frame $1450~\AA$, $M_{1450} = -27$ and $-29$, which set the normalization of the ionizing flux. 

Figure~\ref{figure:constDen} shows a characteristic behavior of a plateau reaching $\rm SB_{\heii}/SB_{\ha} \approx 3-5$ at lower densities, followed by a sharp decline, and a subsequent milder decrease at higher densities.
At lower densities, the clouds are in the optically thin regime and nearly fully ionized both for hydrogen and helium (which in this case, is doubly ionized). 
In this limit, the \heii/\ha\ line ratio approaches a saturation value which is mainly determined by the ratio of the respective, temperature-dependent recombination coefficients. As such, the small variation within this plateau is mainly driven by differences in the cloud temperature, as well as residual variations of the \heiii\ fraction, as determined by the shape and intensity of the incident ionizing flux. 

There is then a second range of densities for which hydrogen remains almost fully ionized while helium becomes more neutral and eventually self-shielded.  
In this regime, the ionization structure becomes stratified. The \hii\ region extends over the whole cloud, while the \heiii\ region is confined to a smaller layer whose \heiii\ column density is fixed and is only dependent on the flux of $E > 54.4$~eV photons. Thus, in this regime, the \ha\ luminosity continues to increase proportional to the density squared, whereas the \heii\ luminosity remains almost constant, with only a modest dependence on temperature. As such, the line ratio drops sharply by $\sim$ one order of magnitude over a factor of $\sim$ 3 in density, after which hydrogen also becomes self-shielded. As expected, the density range within which this transition occurs depends strongly on the quasar flux (i.e., the quasar magnitude and distance of the cloud) and on the cloud size (since the relevant variable for self-shielding is the column density). The hardness of the spectrum changes the relative sizes of the \heiii\ and \hii\ layers, thus changing both the ratio of critical densities for self-shielding of hydrogen and helium and the magnitude of the line ratio drop. 

At higher density, when both hydrogen and helium are self-shielded, the ratio continues to decrease more gradually, reflecting the stabilized stratified structure of the gas and variation in temperature, including contributions from emission produced by collisional excitation in the partially ionized layers. The same radiation field also sets the thermal structure (see Figure~\ref{fig:Tden} in the Appendix), with harder spectra producing more efficient photo-heating.

\subsection{Inhomogeneous Density Distribution}\label{sec:lognorm}

While the single-density scenario explored above is useful to understand the expected behavior of the \heii/\ha\ ratio in a simple case, its applicability to real situations involving extended emission from the CGM is extremely limited. Indeed, the measured fluxes or SB in each spatial resolution element could result from a large number of clouds along our line of sight that could have different properties and that could be spatially located within different CGM regions. 
As demonstrated by \cite{Cantalupo19}, when the gas density distribution of clouds within the spatial aperture is inhomogeneous, the "observed" or measurable line ratio produced by two different ions can be very different from the average "intrinsic" line ratio, i.e. the ratio resulting from the knowledge of local densities in every point in space which has the behavior discussed in the single density scenario. 

Numerous observational studies of hydrogen and helium emission in the CGM \citep{Cantalupo14,Battaia15a,Hennawi15,Cai17b,Cantalupo19} have suggested that broad density distributions are needed in order to explain the properties of the observed emission. 
Unfortunately, these observations cannot constrain the shape of the gas density distribution, which must be assumed or constrained using, e.g., numerical models. 
For instance, \citet{Vossberg19} investigated the resulting gas density distribution produced by cold streams flowing within hot halos that are subject to non-linear processes such as Kelvin-Helmholtz instabilities, finding approximately log-normal density distributions, similar to that produced by turbulent processes shaping the ISM density distribution \citep{Federrath08,Schneider13,Lombardi15}. 
Although current cosmological simulations do not have the spatial resolution of the studies mentioned above, it is interesting to note that they also produce CGM cold gas density distributions consistent with pseudo-log-normals. In particular, by comparing three different cosmological simulations using different numerical models and sub-grid physics, \Laz\ find skewed log-normal PDFs with median densities of order $n \sim 10^{-2}$ and clumping factors $C \sim 15-100$ within the CGM of 10$^{12}$ M$_{\odot}$ halos at $z\sim3$.  

Given the results mentioned above and for consistency with previous analytical models presented in \cite{Cantalupo19}, we adopt the probability function describing a log-normal density distribution, which can be expressed as
\begin{equation} \label{eqPDF}
	PDF[\ln(n)] ~d\ln(n) = \frac{1}{\sqrt{2 \pi} \sigma} \exp \Biggl( -\frac{[\ln(n) - \mu]^2}{2 \sigma^2} \Biggr)  ~d \ln(n),
\end{equation}
where $\mu=\ln(n_0/\rm{cm}^{-3})$ where $n_0$ is the characteristic density that is equivalent to the median density of the distribution in the log-normal case, and $\sigma$ is the log-normal standard deviation (expressed in natural logarithm units), which is related to the clumping factor by the relation $C= e^{\sigma ^2}$. For instance, a $\sigma \simeq 2.628$ corresponds to $C=1000$. Finally, the average volume density ($\langle n \rangle$) is related to the median density by the following relation: $\langle n \rangle=n_0 e^{\sigma^2/2}$. For instance, a distribution with $\sigma \simeq 2.628$, has a mean density about 30 times larger than its median density. However, because our density grid is truncated over the range $10^{-5} \leq n/{\rm cm}^{-3} \leq 10^{5}$, the relations $C=e^{\sigma^2}$ and $\langle n\rangle=n_0e^{\sigma^2/2}$ are only approximate. The corresponding finite-range expressions are provided in Appendix~\ref{appendix:finitegrid}.

In Appendix~\ref {appendix:skewness}, we explore variations on the lognormal distribution. In particular, we implement a skewed log-normal density distribution which has an additional parameter that characterizes high-density tails as found in \Laz. Within the adopted skew-lognormal family, skewness has only a minor effect compared to variations in $\sigma$ at fixed $n_0$, although stronger effects may arise for power-law or multi-component dense tails. We therefore focus on the symmetric log-normal PDF.

\begin{figure}[t]
   \begin{center}
   \includegraphics[height=0.36\textheight,angle=0]{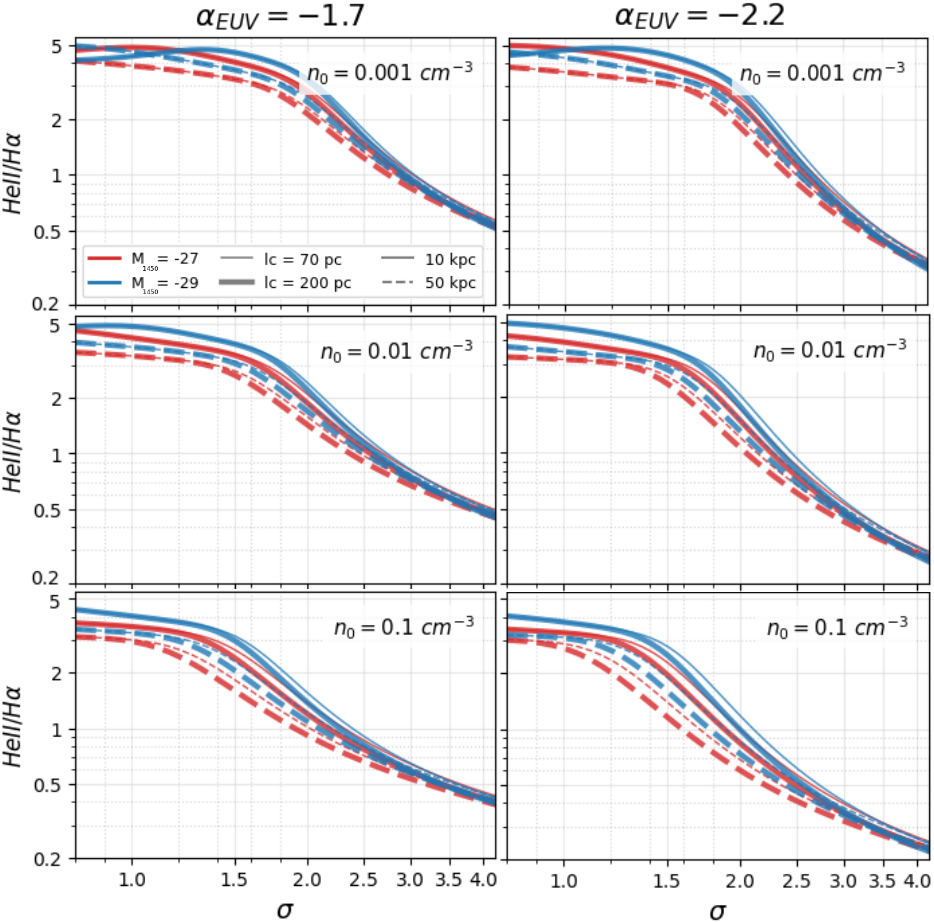} 
   \caption{
   \heii/\ha\ emission-line ratios obtained by our photo-ionization models considering a lognormal PDF density distribution defined by a characteristic density $n_0$, i.e., the median gas density, and standard deviation $\sigma$, which is connected to the PDF clumping factor $C$ by the relation $C=e^{\sigma^2}$. 
   The results are shown for different values of $n_0$ (increasing from top to bottom panels), EUV spectral slopes $\alpha_{\rm EUV}$ (left and right columns), and for different values of $M_{1450}$, cloud distance $D_{\rm c}$, and cloud size $l_c$, as indicated in the legend.}
   \label{figure:lognorm}
   \end{center}
\end{figure}

To derive the PDF-weighted \heii\ and hydrogen line fluxes, we assume a population of clouds at a fixed distance from the quasar with a density distribution given by a log-normal PDF defined by a median density $n_0$, and standard deviation $\sigma$. The cloud distance is kept fixed to isolate the impact of the density PDF from that of the incident ionizing flux. For simplicity, we assume that all clouds have the same size. For each model, we compute the PDF-weighted average luminosity for \ha\ and \heii\ as follows:
\begin{equation}
	L_{line} = \frac{\sum_i L_{line,i} \times PDF(n) ~d ln(n)}{\sum_i  PDF(n) ~d ln(n)}
\end{equation}
where $L_{line}(n)$ is the line luminosity predicted directly by \texttt{CLOUDY} for a cloud of fixed size and therefore already includes the $n^2$ dependence of the line emissivity. 
We explore $n_0 = 10^{-3}$ to $1 ~\rm cm^{-3}$ in 0.5-dex steps, using $10^{-3}$, $10^{-2}$, and $10^{-1}~\rm cm^{-3}$ as the fiducial values shown in Figure~\ref{figure:lognorm} and the higher-density cases as limiting models in Section~\ref{sec:obsmodels}.
The fiducial median density values are motivated by the results of cosmological simulations presented in \Laz, which predict pseudo-log-normal density distributions for the cold CGM in halos of mass $10^{12}$ M$_{\odot}$ at $z\sim3.5$, with median densities in the range $n_0 \sim 10^{-3}-10^{-2}~\rm cm^{-3}$ within spherical shells extending from 10 to 35 physical kpc. These densities are lower than those expected from simple pressure equilibrium with the surrounding hot CGM. For this reason, we also consider a higher median density $n_0 = 0.1~\rm cm^{-3}$, which allows us to explore regimes that may be unresolved in current numerical models, and a lower median density $n_0 = 10^{-3}~\rm cm^{-3}$ for comparison with lower redshift systems. \\

\setlength{\textfloatsep}{-2pt}

Figure~\ref{figure:lognorm} shows the \heii/\ha\ line intensity ratio as a function of $\sigma$, assuming a log-normal distribution. In each plot, different curves represent different combinations of the parameters $l_c$, $M_{1450}$, and $D_c$, as reported in the legend. The left (right) column shows the cases for a harder (softer) AGN SED, $\alpha_{EUV}=-1.7$ (-2.2). Consistent with previous analytical models, our results show that, after an initial plateau, the \heii/\ha\ monotonically decreases with increasing values of $\sigma$, while changing the characteristic density has a smaller effect on the line ratio. 
Moreover, for higher values of $\sigma$ (e.g. $\sigma>2$), the \heii/\ha\ becomes almost independent of the quasar flux as seen from the clouds (i.e., on the combination of quasar luminosity and distance from the quasar) and on the cloud size, at least for the relatively small range of values explored here and relevant for the CGM of bright quasars. 
Instead, the spectral hardness of the ionizing spectrum, parametrized with $\alpha_{EUV}$, is the second most important parameter determining the \heii/\ha\ ratio for large values of $\sigma$.
Specifically, a softer spectrum ($\alpha_{EUV} = -2.2$) leads to a much more rapid decline in \heii\ emission compared to a harder spectrum ($\alpha_{EUV} = -1.7$) as $\sigma$ increases.

\section{Quasar nebulae at $z \simeq 2.2$}\label{sec:z22}

As shown in Section~\ref{sec:lognorm} , the detection and measurement of extended \heii\ and \ha\ within the CGM of quasars has the potential to provide unique constraints on the density distribution of the emitting gas which are free from degeneracies due to different emission mechanisms and radiative transfer effects. In particular, our results demonstrate that observed \heii\ over \ha\ ratios smaller than unity invariably require very broad gas density distributions with log-normal standard deviations $\sigma$ larger than 2, which imply clumping factors $C$ larger than 55 for an infinite density grid. 

In this section, we describe the initial application of our methodology to a sample of Integral-Field and long-slit spectroscopic observations targeting the CGM of bright quasars at $z\approx2.1-2.3$ using KCWI and MOSFIRE for which we can detect both \heii\ and \ha (in addition, of course, to \lya). In Section~\ref{sec:z34}, we extend our model to MUSE observations of the CGM of bright quasars at $z>3.5$ for which, however, \ha\ emission observations are not currently available. 

\subsection{Observations and Data Reduction} \label{sec:ObsData}

Below, we describe the Keck/KCWI and Keck/MOSFIRE observations (PI Reddy; programmes 2025B U239 and U267) of a sample of 13 bright quasars selected at $z\approx2.1-2.3$.
This redshift range was specifically selected to place the \ha\ emission line in relatively clean regions of the near-infrared sky spectrum, minimizing contamination from strong sky lines. The KCWI observations were first used to identify and characterize extended \lya\ and \heii\ nebulae around the quasars (see further information in Table~\ref{tab:KCWI}). Based on the morphology and SB distribution of the detected emission, the MOSFIRE long-slit positions were oriented to intersect the brightest regions of the extended \lya\ and \heii\ emission. Figure~\ref{appfig:slits} shows the MOSFIRE slit positions overlaid on the \lya\ and \heii\ nebulae detected in the KCWI data.\\

The KCWI data were reduced using the standard Data Reduction Pipeline \citep[v1.2][]{Morrissey18}, including bias subtraction, flat-fielding, and flux calibration \citep[more details in][]{Chen21,Song24}. To optimize the reduction for our specific targets, we masked the bright traces of the quasars during the initial stages to prevent the cosmic ray rejection algorithm from misidentifying bright source pixels as artifacts, subsequently improving the PSF subtraction and normalization constants within the \texttt{CubEx} framework \citep{Cantalupo19}. We bypassed the standard KCWI pipeline sky subtraction in favor of the flat-fielding and sky-subtraction refinement using \texttt{CubeKFix} and \texttt{CubeSharp} in \texttt{CubEx}. An initial source mask was generated to identify real astronomical signals, which were then excluded during a second, iterative sky-subtraction procedure to prevent self-subtraction. Individual exposures were spatially aligned through white-light image cross-correlation and flux-normalized to account for atmospheric transparency variations. Finally, the cubes were combined using the \texttt{kcwi\_stack} routine from the KcwiKit suite, where we disabled the \texttt{issue6\_flag}\footnote{see \url{https://github.com/evanhazey/KcwiKit/blob/master/kcwikit/kcwi/kcwi.py} for details.} to bypass a known MontagePy projection bug affecting specific processor architectures and ensure the removal of anomalous negative values. 

The KCWI data were stacked after a systematic quality review of each exposure. For Q1230, despite prevailing poor seeing ($\approx$2.6$\arcsec$), a test-stack using only the three best frames ($\sim$0.8 h) showed no sensitivity gain. We therefore retained the full dataset, which clearly detects both \lya\ and \heii. Similarly, for Q1411, high signal-to-noise detections of both lines were achieved by combining all exposures, despite high airmass and suboptimal seeing ($\approx$2.1$\arcsec$). For J0338-06, excluding frames impacted by passing clouds significantly reduced noise in the central \lya\ region, though \heii\ remains a tentative detection. Conversely, removing individual noisy exposures for J2316+09, J0030+05, and J0010+06 yielded no discernible improvement. Overall, our results rely on the optimal exposure set for each target. Extensive consistency checks confirmed that discarding problematic cubes does not qualitatively alter our findings, except for the improved central \lya\ morphology in J0338-06.\\

The MOSFIRE spectroscopic data were reduced using \texttt{mospy}, the MOSFIRE Data Reduction Pipeline developed by the instrument team and described in \cite{Steidel14}. The observations were obtained in the MOSFIRE mask-nod mode, using interleaved 180 s exposures at two slit positions separated by $\pm20$ arcsec from the fiducial position. For most observations, the central long slit was constructed from 15 aligned masking bars, resulting in a continuous slit approximately 120 arcsec long. The reduction followed the standard multi-slit procedure, including slit-edge tracing and two-dimensional rectification based on the wavelength solution. The K-band wavelength calibration was derived primarily from night-sky emission lines and supplemented by Ne arc spectra at wavelengths longer than $2.3~\mu$m, where the number of usable sky lines is insufficient to robustly constrain the polynomial wavelength solution. Sky subtraction was performed in two stages: first by differencing the A and B frames, and then by fitting the residual background with a two-dimensional B-spline model. Frames affected by clouds were identified through their enhanced thermal background and excluded from the final reduction. 
Each contiguous sequence of nodded exposures was reduced separately. The resulting rectified, wavelength-calibrated, and background-subtracted 2D spectra were shifted to vacuum wavelengths and corrected to the heliocentric reference frame appropriate to the observing date. Spectra obtained during different nights or observing sequences were then combined with the MOSPEC spectral-analysis package \citep{Strom17} using inverse-variance weighting. MOSPEC was also used to perform telluric and flux calibration, producing final 2D spectra in units of flux density per pixel. The observations used 1 arcsec-wide slits, corresponding to a resolving power of approximately $R=2550$ in the K band.

The MOSFIRE data quality was assessed on a target-by-target basis, and observations affected by severe cloud attenuation, unstable seeing, or insufficient integration time were excluded from the analysis (e.g., Q1426+25).
Conversely, October and December 2025 sessions enjoyed clear skies and stable seeing, yielding high-fidelity data with $>2$ hours of integration for J0338-06, J2316+09, and Q0050+00, the latter despite a brief interruption due to high humidity. The SLUG+Rc K-band observations were affected by variable atmospheric transparency. The first frames were taken under cloudy, low-throughput conditions, but the throughput improved during the night and the rest of the sequence was obtained under more stable conditions, with a reported seeing of about 0.5 arcsec.
A summary of the observing dates, exposure times, and seeing conditions for each target is reported in the Table~\ref{tab:MOSFIRE}.

In our analysis, we also include \ha\ emission lines extracted from archival MOSFIRE data, listed in the lower part of Table~\ref{tab:MOSFIRE}. Specifically, we consider a previous slit centered on Q1227+28, the slit targeting J0010+06 presented by \cite{Langen23}, and the slit analyzed by \cite{Leibler18} for the Slug nebula \citep{Cantalupo14}. For Q1227+28, we use unpublished MOSFIRE observations to extract information on the \ha\ emission line at an additional position. For J0010+06 and the Slug nebula, whose MOSFIRE data have already been published, we re-extract the \ha, \heii, and \lya\ emission lines following the same procedure adopted for the rest of our sample, in order to ensure a homogeneous analysis. In these two cases, we also use the \heii\ and \lya\ fluxes reported by \cite{Langen23} and \cite{Cantalupo19}, respectively.

\subsection{PSF/continuum subtraction, and extraction of extended \lya, \heii, and \ha\ emission}\label{sec:anaKeck}

To isolate and extract the extended \lya\ and \heii\ emission from the KCWI data cubes, we employed the \texttt{CubExtractor} package, specifically utilizing the \texttt{CubePSFSub} and \texttt{CubeBKGSub} tools, which are presented in more detail in \cite{Cantalupo19}. For each target, the central quasar's point spread function (PSF) was subtracted using \texttt{CubePSFSub}\footnote{The PSF subtraction of KCWI data with \texttt{CubEx} requires the setting of \texttt{-recenter .false.} in \texttt{CubePSFSub}} by scaling a PSF empirical model constructed from the normalization of pseudo-narrow band defined at each wavelength. Following PSF removal, any residual galactic or quasar continuum was subtracted via \texttt{CubeBKGSub} using a median filtering approach. The extraction of the nebulae was then performed with \texttt{CubEx}, which identifies connected voxels above a certain signal-to-noise ratio ($S/N$) threshold that we range from 2 to 3 per voxel. To ensure the physical validity of the detections, we required a minimum volume for the extracted objects, usually setting \texttt{MinNVox}$>$1000 ($>$100 for the extraction of \heii\ nebulae). This standardized workflow allowed for a robust, reproducible recovery of the faint, extended emission lines surrounding the quasars.

For the MOSFIRE slit spectroscopy, we implemented a custom data reduction workflow specifically designed to isolate faint, extended line emission by accurately removing the bright quasar PSF and the underlying continuum. The procedure consists of three integrated stages. First, a sliding-window PSF subtraction is performed layer-by-layer along the spectral axis. For each wavelength channel, an empirical PSF profile is constructed by collapsing a spectral window of typically 80 pixels. To ensure a clean profile, we carefully exclude regions contaminated by sky lines or the expected astrophysical emission. These problematic regions and instrumental artifacts are identified through an automated multi-scale detection algorithm that scans for coherent vertical structures in the 2D slit using various spatial binning scales (typically between 1 and 40 pixels). The resulting empirical PSF template is then scaled and spatially shifted (allowing for a maximum displacement of 3-5 pixels) to minimize residuals at the quasar position using a robust median-based normalization. Finally, any residual continuum or flat-field variations are corrected using a robust sliding-median estimator along the spectral direction (typically 40-50 pixels).


To ensure a consistent comparison between the different datasets and estimate the line ratios between \heii, \lya\ (KCWI data), and \ha\ (MOSFIRE data) emission lines in the CGM, we converted the KCWI datacubes into synthetic 2D slit spectra. Since all the MOSFIRE slits are centered on the quasars, this common reference point allowed for a precise relative WCS alignment and slit positioning.
To perform this conversion, we extracted the flux from the 3D datacubes within a spatial mask reproducing the geometry of each MOSFIRE slit \footnote{If local resonant scattering is important, photons may be redistributed outside the slit aperture, and ratios with \lya\ in the denominator should therefore be interpreted as upper limits.}. The slit orientation was determined using the \texttt{ROTPOSN} keyword together with the sky coordinates of the masked pixels (\texttt{TARGRA}, \texttt{TARGDEC}). 
The spectra of all pixels enclosed within the slit mask were then projected onto the principal slit axis. The final 2D slit-like images were constructed by collapsing the flux distribution along the direction perpendicular to the slit and computing the mean spectral SB within each spatial bin. This approach preserves the average SB across the slit width while removing any dependence on the native pixel scale of the two instruments, thereby enabling a direct comparison between the KCWI-derived slit images and the MOSFIRE 2D spectra.
This procedure collapses the 3D information into a 2D spectral-spatial format, matching the resolution and orientation of the corresponding slit observations and enabling a direct, pixel-to-pixel comparison of the different emission lines. \\

\subsection{Line ratios within 35 kpc from the quasars}\label{sec:ratiosz2p2}

To measure the emission-line fluxes and the corresponding line ratios, we developed a dedicated interactive pipeline to perform spatially integrated spectroscopy on the two-dimensional slit spectra. As a first step, all datasets were homogenized onto a common flux scale. In particular, the synthetic KCWI slits were rescaled by accounting for both the spatial pixel size and the spectral dispersion, to ensure consistency with the MOSFIRE flux-per-pixel calibration. Throughout this analysis, fluxes are expressed in units of $10^{-19}~{\rm erg~s^{-1}~cm^{-2}}$.

The flux extraction was performed through a supervised interactive interface, which allowed us to define the spatial and spectral integration regions for each target and line. We adopted a fiducial maximum projected spatial extent from $\approx$5-10 to 35 pkpc from the quasar position. As a robustness check, we repeated the measurements using apertures of 30, 35, and 50 pkpc, finding consistent fluxes and line ratios within $1\sigma$. To ensure a homogeneous comparison among the three transitions, the same spatial mask was applied to the \ha, \lya, and \heii\ slit spectra. In particular, we excluded the central region affected by quasar PSF residuals, identified by strong positive and negative flux excursions. This conservative masking strategy ensured that the three line fluxes are measured over the same physical region.
A key step of the pipeline is the interactive correction of the local background, performed both spatially and spectrally. Along the slit, local background regions were selected to estimate the residual background while avoiding contamination from extended nebular emission, sky-line residuals, or instrumental artifacts. In addition, the zero level of the extracted one-dimensional spectrum was manually adjusted before flux integration to remove residual continuum or sky offsets. This combined spatial and spectral continuum and background correction accounts for both large-scale background gradients and small residual offsets that could otherwise affect the final line-flux measurements.

\begin{figure}[t]
   \begin{center}
   \includegraphics[height=0.5\textheight,angle=0]{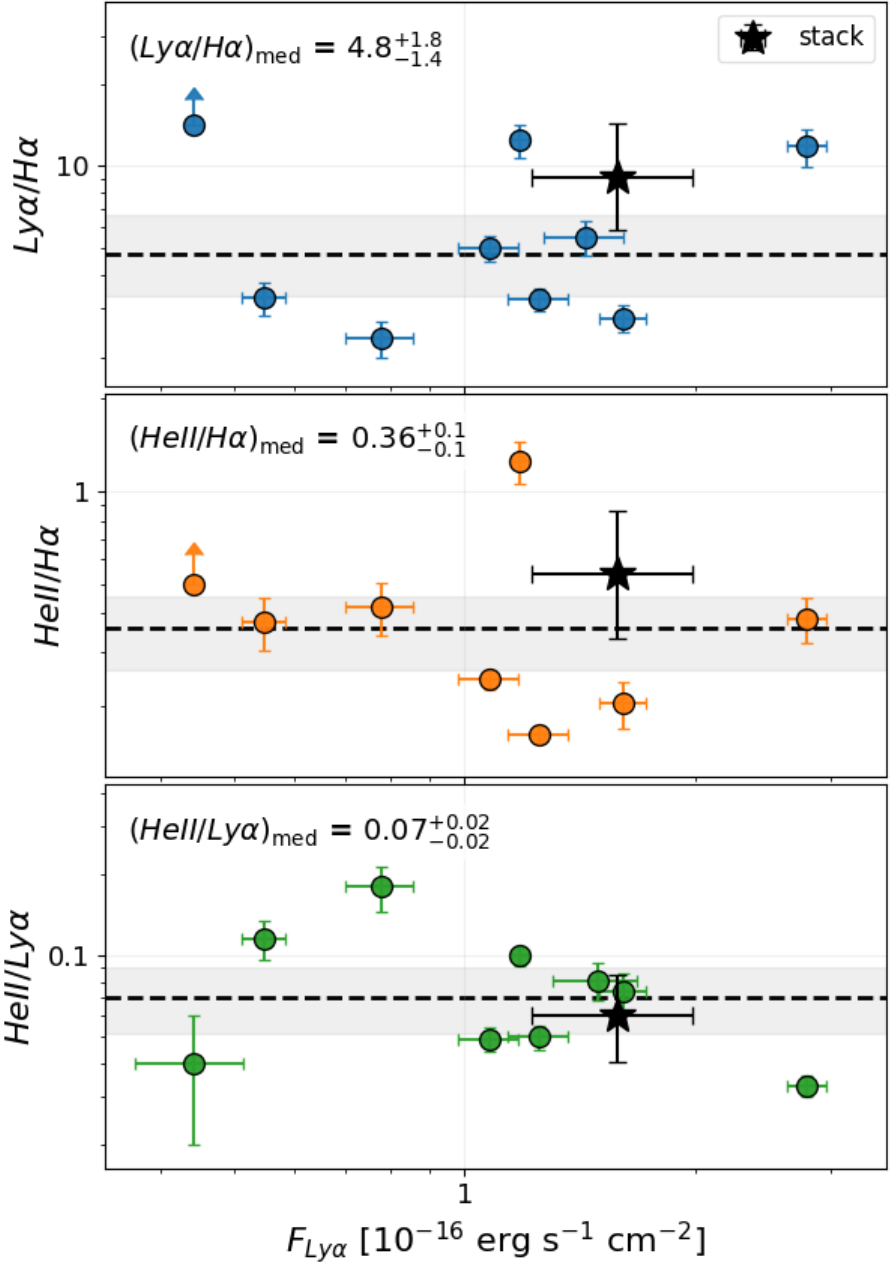} 
   \caption{Line-ratios within the area defined by the MOSFIRE slits (see Figure~\ref{appfigure:CGMHaLyaHe}) and within a distance of $\approx$35 kpc from the quasars for the nebular emission in our sample at $z\sim2.3$. From top to bottom, we show \lya/\ha, \heii/\ha, and \heii/\lya\ as a function of the integrated \lya\ flux, measured within the MOSFIRE and KCWI common spatial aperture after masking the central quasar PSF residuals and integrating out to 35 kpc. Dashed lines and grey bands indicate the median ratios and 16th-84th percentile ranges estimated through our Monte Carlo procedure, accounting for measurement errors and finite-sample variance. The black stars show the ratios measured from the stacked spectra. See text for details.}

   \label{figure:HeHaLya1}
   \end{center}
\end{figure}
\setlength{\textfloatsep}{-2pt}

To obtain robust flux estimates, we repeated the extraction for each slit multiple times, varying the adopted spectral window and the size of the central PSF-masked region. In each realization, the integration window was chosen to integrate all the line emission visible in the extracted spectrum while avoiding contamination from sky-line residuals and cosmic rays. In some cases, we adopted fixed windows of approximately $\pm500~{\rm km~s^{-1}}$ for \ha\ and \heii, and broader windows of $\pm1000$--$1200~{\rm km~s^{-1}}$ for \lya, reflecting the broader kinematic extent of the resonant \lya\ emission. In cases of non-detection, we estimated upper limits at the $2\sigma$ level. The final flux of each line was obtained as the average over the independent extractions. The associated uncertainty combines the propagated measurement errors and the scatter among the different realizations. For each object, line ratios and their uncertainties were then computed from these averaged fluxes through standard error propagation. The median line ratios of the sample were instead estimated through a Monte Carlo procedure that accounts for measurement uncertainties and finite-sample variance. This conservative strategy reduces the dependence of the results on any single interactive choice and provides robust flux, line-ratio, and sample-median estimates.

We applied this procedure to the targets for which reliable \ha, \lya, and/or \heii\ measurements could be obtained. For the computation of the median line ratios shown in Table~\ref{tab:final_line_ratios} and Figure~\ref{figure:HeHaLya1}, we adopted the following selection. 
J1426+25 was observed for only 25 min under cloudy, low-throughput conditions, making its extracted \ha\ flux unreliable. We therefore exclude this object from the \lya/\ha\ and \heii/\ha\ diagnostics, while retaining it for \heii/\lya, which relies only on the KCWI data. J2316+09 is excluded from the final sample because its \lya\ and \heii\ nebulae are only marginally extended, the observations were affected by poor weather, and the MOSFIRE data do not provide a robust \ha\ constraint. Conversely, we retain CH-Q02, although it is 1.5-2 magnitudes fainter than the average quasar luminosity of the sample (see Table~\ref{tab:KCWI}, col. $mag~i$), since our lognormal-density photo-ionization models show that variations of the ionizing luminosity by up to two magnitudes have only a limited impact on the predicted \heii/\ha\ ratio (Figure~\ref{figure:lognorm}). We also include J0010+06, using the MOSFIRE data presented in \cite{Langen23} together with our KCWI \lya\ and \heii\ measurements. For Q1227+28, we use the average fluxes obtained from the three available MOSFIRE slit measurements. Finally, we include the SLUG \lya/\ha\ measurement from \cite{Leibler18} as a literature comparison point. For SLUG-Rc, the MOSFIRE slit is centered on source C, while the corresponding \lya\ and \heii\ measurements are taken from \cite{Cantalupo19}.\\

As a complementary measurement, we constructed stacked one-dimensional spectra for the three lines. Starting from a fiducial interactive extraction, we generated multiple randomized realizations by varying the extraction aperture, PSF mask, masked spectral/spatial regions for sky lines and spectral features, background level, velocity centering, and integration windows within the adopted ranges $\pm 1000~\rm km~s^{-1}$. For each realization, the spectra were shifted to the systemic frame, continuum-zero subtracted, stacked, and integrated over the selected velocity windows. The stacked measurements are shown as black stars in Figure~\ref{figure:HeHaLya1} and reported in Table~\ref{tab:final_line_ratios}. The resulting stacked line ratios are $\lya/\ha = 9.05_{-3.24}^{+5.30}$, $\heii/\ha = 0.54_{-0.21}^{+0.32}$, and $\heii/\lya = 0.060_{-0.020}^{+0.025}$. These values are consistent, within 1$\sigma$ uncertainties, with the object-by-object median estimates. 

The median of the individual measurements yields $\lya/\ha \simeq 5 \pm 2$, which is consistent with the case B (case A) recombination value of 8.3 (12.7) at a temperature of $2\times10^4$ K once the resonant redistribution of \lya\ photons outside the slit is taken into account (see \citet{Leibler18,Langen23} for details). In contrast, the stacked method, which also includes sources with upper limits on \heii, gives a higher ratio, $\lya/\ha > 6$ within $1\sigma$. Collisional excitation or resonant photon pumping, which are expected to produce significantly larger \lya/\ha\ ratios \citep[>100; see][]{Langen23,Ledos26}, do not dominate the \lya\ emission within the regions sampled by our MOSFIRE observations. However, since the MOSFIRE slits were placed on the brightest \lya\ and \heii\ emitting regions, the measured ratio does not necessarily represent the total \lya\ photon budget of the nebula, which may be redistributed over larger scales by resonant scattering.

\begin{figure}[t]
   \begin{center}
   \includegraphics[height=0.36\textheight,angle=0]{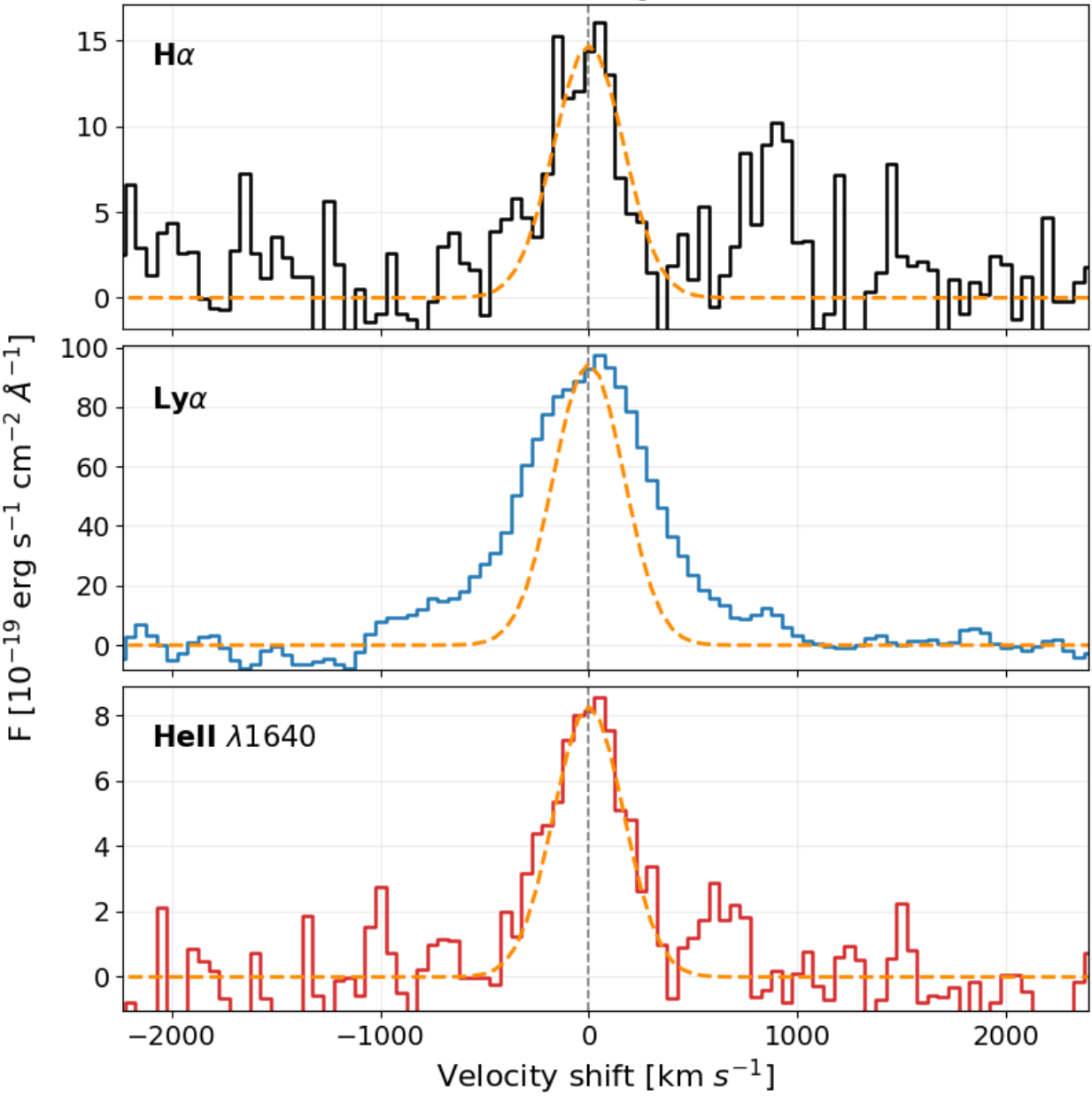} 
   \caption{Average stacked spectra of spatial regions contained within the MOSFIRE slits and within a distance of 35 kpc from the quasars for our sample at $z\sim2.3$. The x-axis shows velocity in ${\rm km~s^{-1}}$ relative to the line peak. The dashed orange curves show the best-fit Gaussian profiles of the stacked \ha\ emission line.}
   \label{figure:spectra}
   \end{center}
\end{figure}
\setlength{\textfloatsep}{-2pt}
In Figure~\ref{figure:spectra}, we show the stacked spectra of \ha, \lya, and \heii. The \ha\ and \heii\ profiles are well described, to first order, by single Gaussian functions centered at the systemic velocity, and we use these fits to estimate their velocity dispersions, $\sigma_{\ha}$ and $\sigma_{\heii}$. We do not fit the \lya\ profile with a Gaussian, because resonant scattering can produce asymmetric, non-Gaussian line shapes. Instead, we measure the FWHM of the stacked \lya\ profile directly and define an equivalent Gaussian dispersion as $\sigma_{\lya, \rm eq} = \rm{FWHM}_{\lya}/2.355$. This quantity should not be interpreted as evidence that the \lya\ profile is Gaussian, but provides a uniform width estimator for comparison with the non-resonant lines. The widths measured from the stacked spectra may be broadened by residual velocity-alignment uncertainties and object-to-object variations, and should therefore be regarded as upper limits to the typical intrinsic dispersions. Nevertheless, their relative values remain useful for comparing the kinematics traced by the different transitions. For comparison, we overlaid the best-fit Gaussian profile of the \ha\ emission onto the \lya\ and \heii\ panels as a dashed orange line, to directly contrast their widths.

From the Gaussian fits to the stacked non-resonant lines, after correcting for the instrumental line spread function (LSF\footnote{We adopt instrumental FWHM values of $118~\rm km~s^{-1}$ for MOSFIRE in the K-band (R$\simeq 2250$), and $187~\rm km~s^{-1}$ or KCWI with the BL grating and Medium slicer (R$\simeq 1600$), corresponding to Gaussian dispersions of $\sim 50$ and 80$\rm ~km~s^{-1}$, respectively.}), we obtain $\sigma_{\ha} = 165^{+31}_{-27}~{\rm km~s^{-1}}$ and $\sigma_{\heii} = 177^{+22}_{-20}~{\rm km~s^{-1}}$. For \lya, we measure the FWHM of the stacked profile and convert it into an equivalent Gaussian dispersion, obtaining $\sigma_{\lya} = 295^{+35}_{-40}~{\rm km~s^{-1}}$. The resulting width ratios are $\sigma_{\lya}/\sigma_{\ha} = 1.79^{+0.60}_{-0.49}$, $\sigma_{\heii}/\sigma_{\ha} = 1.07^{+0.37}_{-0.27}$, and $\sigma_{\lya}/\sigma_{\heii} = 1.67^{+0.44}_{-0.39}$. As expected, the two non-resonant lines have comparable widths, with their ratio being consistent with unity. In contrast, the resonant \lya\ line is broader than the non-resonant lines by a factor of approximately $1.3-2.4$, consistent with the additional broadening expected from resonant scattering and radiative-transfer effects.

\section{Quasar nebulae at $3.1 <z<4.5$} \label{sec:z34}

MUSE observations of quasars at $z>3$ have revealed the ubiquitous presence of extended and bright \lya\ emission on CGM scales \citet{Borisova16,Battaia19,Farina19,Muzahid20,Fossati21}, providing a large statistical sample to explore the evolution of the CGM properties over time. Unfortunately, the redshift range of these observations does not allow the detection of \ha\ emission from the ground. Nonetheless, we will take advantage of our observational results at $z\sim2.2$ presented in Section~\ref{sec:z22} to provide new constraints on the CGM of MUSE observations around bright quasars.  

\subsection{Observations and Data Reduction of MUSE data} \label{sec:ObsData}

To detect and analyze \lya\ and \heii\ emission surrounding quasars in the redshift range of $3.1 < z < 4.5$, we selected MUSE observations with an exposure time of at least 4 hours on source. We considered this exposure time as a minimum to detect potentially faint \heii\ emission. 
The entire sample we analyzed consists of 26 quasars already presented in other works \citep[e.g.,][see Table~\ref{tabapp:magg}]{Borisova16,Lofthouse20,Fossati21}. Most of these observations were carried out as part of the MUSE Large Program \citep[ID 197.A-0384, PI Fumagalli;][MAGG sample]{Lofthouse20}, and are supplemented by data from the MUSE GTO \citep[PI J. Schaye;][MUSEQuBES sample]{Muzahid20}. These data were reduced by applying a bias subtraction and correcting for flat-fielding, twilight, and illumination on the raw MUSE data, using the standard ESO MUSE pipeline \citep[v2.4.1][]{Weilbacher14}. Afterwards, a more accurate flat-fielding correction and the sky subtraction were performed with the \texttt{CubeFix} and \texttt{CubeSharp} tools. More details about these tools belonging to the \texttt{CubExtractor} package are in \cite{Borisova16,Cantalupo19}.

\subsection{Detection of \lya\ and \heii\ nebulae} \label{sec:method}

As described in Section~\ref{sec:anaKeck}, we extract the \lya\ and \heii\ nebulae by following the iterative PSF-subtraction procedure described in \cite{Borisova16} and \cite{Cantalupo19}. We used the \texttt{CubePSFSub} tool to generate an empirical PSF model, assuming the central pixels are dominated by quasar emission. Initially, a standard spectral filter width of 150 layers was applied. After a visual inspection to identify the spectral regions containing \lya\ and \heii\ emission, we performed a second run using a customized PSF and masking the expected extended transitions to prevent signal self-subtraction. To optimize the subtraction, we tested various filter widths (80 to 200 layers), selecting the one that minimized the flux standard deviation within a 5-pixel diameter circle at the PSF center. Background and continuum subtraction for other sources in the field were performed using the \texttt{CubeBKGSub} tool via median filtering. We then used \texttt{CubeSel} to extract sub-cubes centered on the lines of interest within a range of $\pm 100$ layers ($\sim 5500 \, \rm km \, s^{-1}$ at $7000 \,\AA$). The final extraction of the nebulae was performed with the \texttt{CubEx} tool \citep{Cantalupo19}. We applied spatial and spectral smoothing using a Gaussian kernel (radius 0.4$\arcsec$ and 1.25~$\AA$, respectively). Detection thresholds were set to a signal-to-noise ratio (S/N) of 3 (\lya) and 2.5 (\heii), requiring a minimum of 8000 and 1000 connected voxels, respectively. Sky lines were identified via a 3$\sigma$ sigma-clipping algorithm and masked. This approach returns a \texttt{3D-mask} that serves to define the nebula region, enclosing all connected voxels above a certain signal-to-noise threshold.
While \lya\ nebulae were detected around all quasars, \heii\ emission was confirmed in 14 out of 26 cases ($54\%$ of the sample), at a median redshift of $z_{\rm med} \simeq  3.671$ derived from the \heii\ lines. The resulting SB maps are shown in Figure~\ref{figure:SBmaps}. The reliability of the fainter \heii\ detections was further verified through statistical tests (see Appendix~\ref{app:spec}). Uncertainties (1$\sigma$) for SB, flux, and luminosity were estimated by flux-resampling. We applied 100 random wavelength shifts to the 3D masks and calculated the standard deviation of the resulting flux distributions. For radial profiles (SB and velocity dispersion), uncertainties correspond to the 25th and 75th percentiles derived from 500 bootstrap iterations.
The stacked radial SB profiles of \lya\ and \heii\ from the combined KCWI and MUSE samples show very similar shapes, with ${\rm SB}_{\rm HeII}/{\rm SB}_{\rm Ly\alpha} \simeq 0.05$ remaining approximately constant out to $\sim150-160$~ckpc (Figure~\ref{figure:RatioHeIILyaSB}). This demonstrates that our assumption on the co-spatial nature of \heii\ and hydrogen emission is well justified.

\subsection{\heii/\lya\ line ratios of CGM in $3.1 < z < 4.5$ quasars} \label{sec:ratios}

\begin{figure}[t]
   \begin{center}
   \includegraphics[height=0.22\textheight,angle=0]{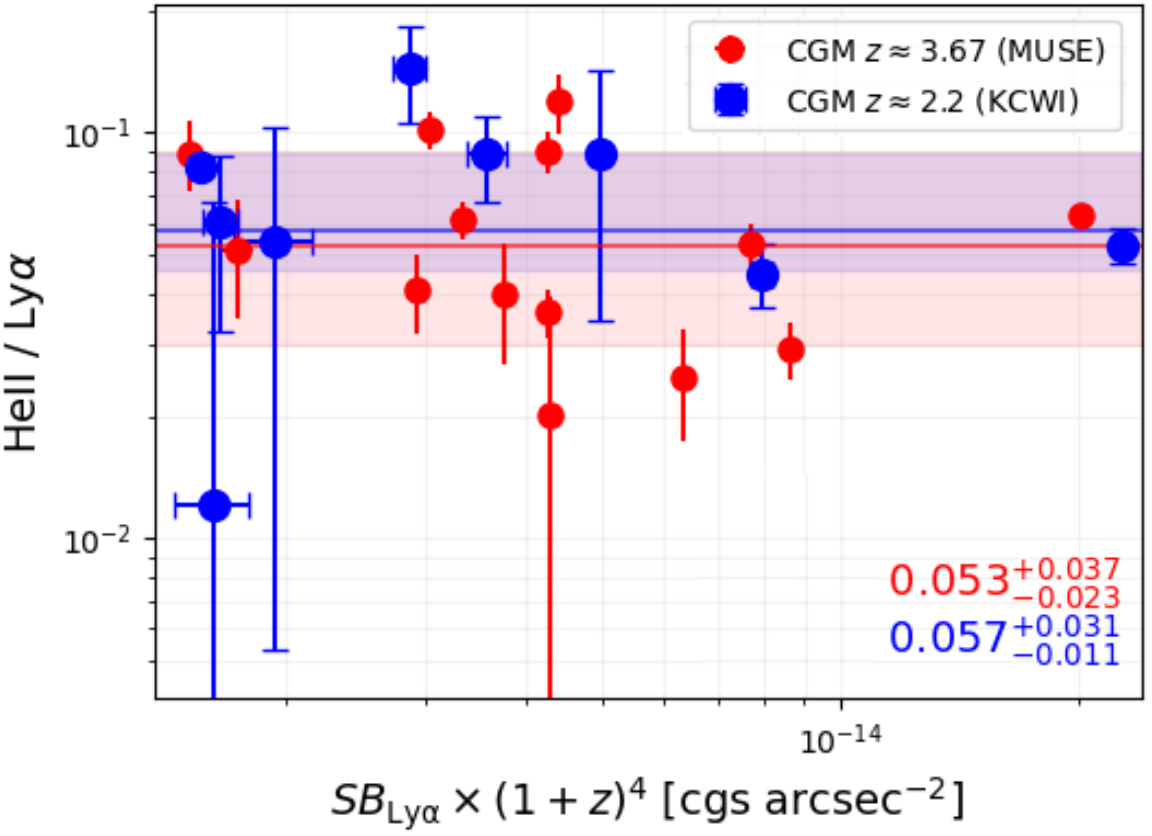}   
   \caption{Line ratios as a function of the redshift-dimming corrected $SB_{\lya}$ emission of each nebula in our total sample at different redshifts. Unlike Figures~\ref{figure:HeHaLya1} and \ref{figure:spectra}, the area considered here is the region contained within an annulus with internal and external radii of 50 and 160~$\rm ckpc$, respectively. The horizontal solid lines, along with the transparent bands, show the median and 16th/84th percentiles \heii/\lya\ line ratios for the detected emission at z$\approx2.2$ (blue) and $z \approx 3.67$ (red).}
   \label{figure:SBratio}
   \end{center}
\end{figure}

Figure~\ref{figure:SBratio} shows the cosmological dimming corrected $\rm SB_{\lya}$ versus the $\rm SB_{\heii}/SB_{\lya}$ (named \heii/\lya\ because equivalent to  $\langle F_{\heii} \rangle / \langle F_{\lya} \rangle$ since extracted from the same area) line-ratio. We calculated the \heii/\lya\ emission line ratios for the individual nebulae within an annulus ranging from 50 to 160~$\rm ckpc$\footnote{As for the annular regions adopted for the KCWI and MOSFIRE observations, this annulus was chosen to exclude the inner region affected by PSF-subtraction residuals and the outer region where the \heii\ emission approaches the surface-brightness detection limit.}. The emission of the nebulae was derived in pseudo-narrow band images obtained by collapsing the datacube within the spectral range defined by the \texttt{3D-mask} and by removing spectral regions dominated by residual sky lines. 
The uncertainties shown in the plots represent the 1$\sigma$ errors estimated using the procedure described in Section~\ref{sec:method}. 
Red dots show the \heii/\lya\ values for the sample of CGM-line emitting nebulae around quasars at $z_{\rm med} \simeq  3.671$. For these cases, the 50-160~$\rm ckpc$ annulus corresponds to 10.5-33.6~$\rm pkpc$. 
The same methodology described above is applied to estimate the \heii/\lya\ line ratio within the 50-160~ckpc annulus, for the CGM systems at $z_{\rm med} \simeq 2.2$, in the KCWI dataset (Table~\ref{tab:KCWI}), for the cases in which both \heii\ and \lya\ nebulae are detected. These measurements are shown as blue dots in Figure~\ref{figure:SBratio}, allowing a direct comparison between the CGM at different redshifts.

\setlength{\textfloatsep}{-2pt}

We note that, for the CGM at $z_{\rm med}  \simeq 2.2$ and $z_{\rm med} \simeq 3.671$ the median values of \heii/\lya\ are in agreement, i.e., $\approx 0.05_{-0.02}^{+0.04}$, marked by the union of transparent blue and red bands representing the 16th and 84th percentiles of the line ratios of the sample. 
Important assumptions that we made are that the \lya\ and \heii\ emitting clouds are co-spatial and share the same kinematics. To assess the impact of this assumption on our results, we derived the $\rm SB_{\lya}$ values from pseudo-NB images obtained by centering a top-hat filter at the redshift derived from the \heii\ emission (i.e., $z_{\heii}$). We tested the impact of the adopted spectral window by considering widths equal to 1, 2, and 4 times that of the \heii\ nebula mask. The choice of spectral range has a negligible effect on the median value, with variations up to $\sim 3\%$ relative to the fiducial estimate. In all cases, the resulting $\rm SB_{\heii}/SB_{\lya}$ ratios remain below 0.23 (case A) and 0.3 (case B), consistent with the analytical expectations for fully photoionized gas. 
A more detailed discussion of the upper limits derived for systems without detected \heii\ nebulae, together with the implications for the detectability of faint and diffuse \heii\ emission, is presented in Appendix~\ref{app:noheii}.

\subsection{Velocity-width comparison between \lya\ and \heii\ at different redshifts}

We applied the same stacking procedure to the \lya\ and \heii\ emission associated with the sample of nebulae at $z=3.1-4.5$ (median $z = 3.67$). For each object, we extract the spectra from the nebular mask within an annulus of 50-160 ckpc from the quasar position, shift the spectra to a common velocity frame, and stack them. As for the lower-redshift KCWI stack, we fit the non-resonant \heii\ line with a single Gaussian and estimated the \lya\ width through the equivalent dispersion. After correcting for the wavelength-dependent MUSE LSF, adopting $FWHM \sim 2.44$ and $2.71~ \AA$ at the observed wavelengths of \heii\ and \lya, respectively, we obtain $\sigma_{\heii}=197^{+20}_{-20}~{\rm km~s^{-1}}$ and $\sigma_{\lya,{\rm eq}}=504^{+34}_{-58}~{\rm km~s^{-1}}$. These values correspond to $\sigma_{\lya}/\sigma_{\heii}=2.56^{+0.49}_{-0.51}$.

This ratio is larger than that measured for the sample of nebulae at $z_{\rm med} \simeq 2.2$, where we found $\sigma_{\lya}/\sigma_{\heii}=1.67^{+0.44}_{-0.39}$. The \heii\ widths are therefore broadly comparable in the two stacks, while the \lya\ profile appears significantly broader in the higher-z sample. This may indicate stronger resonant broadening or radiative-transfer effects at higher redshift. However, we caution that the comparison is not purely evolutionary. 
An additional caveat is that the two measurements probe different regions of the CGM. In the KCWI/MOSFIRE sample, the slit positions were selected to maximize the expected \heii\ signal and therefore preferentially target the brightest and most highly ionized structures. By contrast, the MUSE measurement includes all nebular emission within the adopted annulus. Consistent with this selection effect, the integrated \heii/\lya\ ratio decreases when measured over the full annulus rather than within the slit regions. If the brightest \heii\ regions also correspond to lower neutral-hydrogen optical depths, they may be less affected by resonant scattering, potentially reducing the observed \lya\ broadening. Therefore, part of the difference in $\sigma_{\lya}/\sigma_{\heii}$ between the two samples may reflect the different spatial selection rather than a genuine redshift evolution.
Moreover, because for the higher-z sample we extracted the spectrum from an extended annulus out to 160 ckpc, we may include a larger range of spatially varying velocities, scattering paths, and centroid uncertainties. Therefore, while the higher-redshift stack suggests a larger relative broadening of \lya\ with respect to \heii, we interpret this result as indicative rather than conclusive.

\section{Constraining the CGM density distribution}\label{sec:obsmodels}

\begin{figure}[t]
   \begin{center}
   \includegraphics[height=0.44\textheight,angle=0]{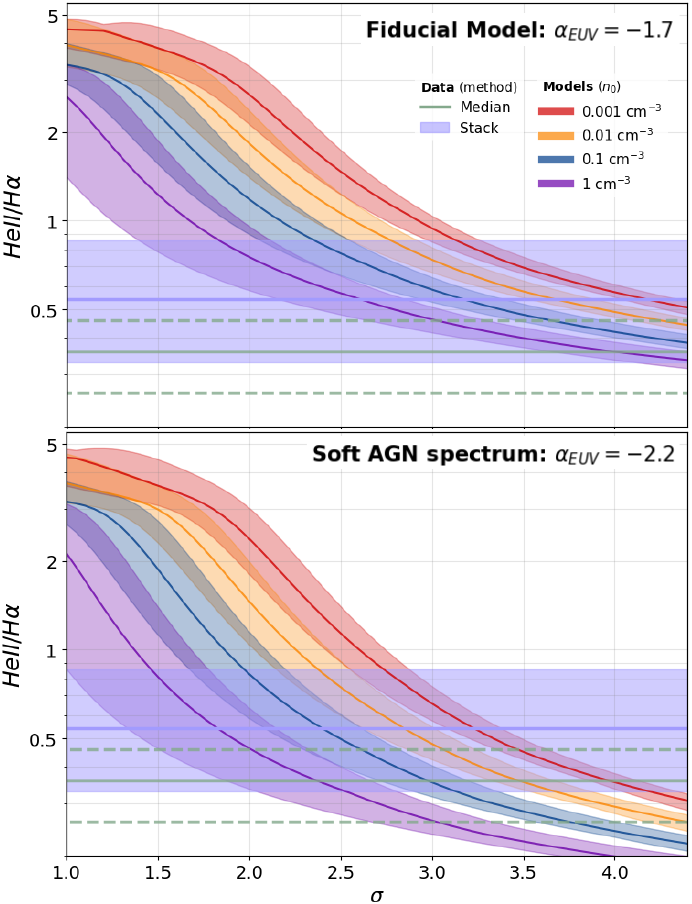}
  \caption{\heii/\ha\ emission line ratio as a function of $\sigma$ and $C$. The panels illustrate \texttt{CLOUDY} model predictions assuming an AGN SED with $\alpha_{\rm EUV} = -1.7$ (top; fiducial model) and $-2.2$ (bottom), assuming a log-normal density distribution. Colored bands represent the full range of \texttt{CLOUDY} models for different characteristic densities: $n_0 = 10^{-3}\,\text{cm}^{-3}$ (red), $10^{-2}\,\text{cm}^{-3}$ (orange), $10^{-1}\,\text{cm}^{-3}$ (blue), and $1 \, \text{cm}^{-3}$ (purple); the width of each band encompasses the variance introduced by $M_{1450}$, $D_c$, and $l_c$. The lavender horizontal shaded area indicates the observed CGM ratio obtained from the stacked spectra. The green horizontal solid and dashed lines mark the median and 16th-84th percentiles of the sample of individual \heii/\ha\ ratios, respectively.}
   \label{figure:Clump}
   \end{center}
\end{figure}

In Sections~\ref{sec:ratiosz2p2}, we measured the \heii/\ha, \heii/\lya, and \lya/\ha\ line ratios in the CGM of quasars at $z\simeq2.2$. For the KCWI-MOSFIRE sample, the stacked spectra yield $\lya/\ha \simeq 9.05_{-3.24}^{+5.30}$ and $\heii/\ha=0.54_{-0.21}^{+0.32}$ within the MOSFIRE slit regions, within a maximum projected distance of $\sim35$ pkpc. 
The consistent \heii/\lya\ ratios measured in the KCWI and MUSE samples at $z \sim 2.2$ and $3<z<4.5$, respectively, show no evidence for strong evolution in the detected, emission-weighted component over the redshift range considered (see Section~\ref{sec:z34}). 
Single-density photoionization models would require $n_{\rm H}\gtrsim 10~\rm cm^{-3}$ to reproduce the stacked \heii/\ha\ ratio. Such densities are difficult to interpret as a volume-filling CGM phase in long-term pressure equilibrium \citep[see][]{Pezzulli19,deBeer23}. 
Indeed, even without considering the photo-heating effect of the quasar (which decreases the density contrast with the hot halo by increasing the cold gas temperature), pressure-equilibrium analytical considerations would imply cold gas densities in the maximum range $0.5-5$ cm$^{-3}$ within 30 kpc of a 10$^{12}$ M$_{\odot}$ halo at $z\sim2.2$ \citep{Aung24,Ledos26}. This instead motivates a scenario in which the recombination emission arises from a broad distribution of gas densities rather than a single characteristic density. We therefore compare the observations with \texttt{CLOUDY} models in which the gas density follows a log-normal distribution. Figure~\ref{figure:Clump} shows the predicted \heii/\ha\ ratio as a function of the width of the density distribution and the corresponding clumping factor. The lavender horizontal shaded area shows the stacked \heii/\ha\ constraint and its 16th-84th percentile range. Because the \texttt{CLOUDY} grid covers only the finite density range $10^{-5}-10^{5}\ {\rm cm^{-3}}$, the inferred clumping factors are those of truncated log-normal PDFs rather than distributions extending over $0<n<\infty$ (Appendix~\ref{appendix:finitegrid}). Consequently, once the PDF becomes sufficiently broad, truncation removes part of the high-density tail that dominates $\langle n^2\rangle$, causing the inferred clumping factor to depart from $C=\exp(\sigma^2)$ and eventually decrease with increasing $\sigma$.
\setlength{\textfloatsep}{-2pt}
For the fiducial ionizing spectrum, $\alpha_{\rm EUV}=-1.7$, the stacked ratio requires $\sigma \gtrsim 3$, 2.5, and 2.1 for the fiducial densities $n_0=10^{-3}$, $10^{-2}$, and $10^{-1}~\rm cm^{-3}$, respectively (red, orange, and blue curves in Figure~\ref{figure:Clump}). The models reproducing the median ratio have $\sigma \simeq 4.2$, 3.6, and 3.2, corresponding to $7\times10^2\lesssim C\lesssim4\times10^3$. An ultra-soft spectrum, $\alpha_{\rm EUV}=-2.2$, reduces the required density dispersion: for the same values of $n_0$, the lower limits become $\sigma_{\rm min}\simeq2.5$, 2.1, and 1.6, while the models reproducing the median ratio have $\sigma\simeq3.2$, 2.8, and 2.4. The inferred lower limits and median constraints on $\sigma$, $C$, and the average volume density $\langle n\rangle$ are summarized in Table~\ref{tab:modelobsC}.
Although $n_0=1~\rm cm^{-3}$ (purple curves in Figure~\ref{figure:Clump}) lowers the required dispersion to $\sigma \approx 2.6$, the corresponding lognormal has $\langle n \rangle \sim 29~\rm cm^{-3}$ ($C \approx 1000$) and places half of its volume above $1~\rm cm^{-3}$. It therefore shifts the tension from the width of the PDF to its overall density and is better regarded as a limiting case than as a preferred physical solution. 

\begin{table}
\centering
\caption{Constraints on the properties of the log-normal density distribution inferred from the comparison between the stacked \heii/\ha\ line ratio and \texttt{CLOUDY} models. For each model, $\sigma_{min}$, $C_{min}$, and $\langle n \rangle_{\rm min}$, indicate the minimum values required for the model to reach the upper $1 \sigma$ statistical limit of the observed stacked ratio. The last column reports the values of $\sigma$, $C$, and $\langle n \rangle$ at which the model prediction, including its $1\sigma$ statistical range, intersects the median observed stacked ratio. The last three columns report the values of $\sigma$, $C$, and $\langle n\rangle$ at which the model prediction, including its $1\sigma$ statistical range, intersects the median observed stacked ratio.}
\label{tab:modelobsC}
\scriptsize
\setlength{\tabcolsep}{3pt}
\resizebox{\columnwidth}{!}{%
\begin{tabular}{c|cccccc}
\hline
\noalign{\smallskip}
$\log\left(\frac{n_{0}}{\rm cm^{-3}}\right)$ & $\sigma_{\rm min}$ & $\log_{10}(C_{\rm min})$ & $\langle n\rangle_{\rm min}$ & $\sigma$ & $\log_{10}C$ & $\langle n\rangle$ \\
 &  &  & $\rm [cm^{-3}]$ &  &  & $\rm [cm^{-3}]$ \\
\noalign{\smallskip}
\hline
\multicolumn{7}{c}{Fiducial models ($\alpha_{\rm EUV}=-1.7$)} \\
\hline
$-3$   & $3.0$ & $3.63$ & 0.10 & $4.2_{-0.2} ^{+0.1}$ & $3.56_{-0.08}^{+0.15}$ & $4.50_{-2.02}^{+1.46}$ \\
$-2.5$ & $2.7$ & $3.08$ & 0.12 & $3.9_{-0.2} ^{+0.2}$ & $3.45_{-0.14}^{+0.12}$ & $4.79_{-2.16}^{+3.58}$  \\
$-2$   & $2.5$ & $2.68$ & 0.23 & $3.6_{-0.2} ^{+0.2}$ & $3.31_{-0.11}^{+0.08}$ & $5.43_{-2.42}^{+4.06}$ \\
$-1.5$ & $2.3$ & $2.29$ & 0.45 & $3.4_{-0.2} ^{+0.2}$ & $3.08_{-0.09}^{+0.05}$ & $8.70_{-3.75}^{+6.16}$ \\
$-1$   & $2.1$ & $1.91$ & 0.91 & $3.2_{-0.2} ^{+0.2}$ & $2.84_{-0.08}^{+0.03}$ & $14.56_{-6.03}^{+9.69}$ \\
$-0.5$ & $1.7$ & $1.26$ & 1.34 & $2.8_{-0.3} ^{+0.2}$ & $2.59_{-0.15}^{+0.01}$ & $15.26_{-8.10}^{+10.05}$ \\
$0$    & $1.4$ & $0.85$ & 2.66 & $2.6_{-0.4} ^{+0.2}$ & $2.31_{-0.30}^{+0.02}$ & $28.38_{-17.15}^{+17.25}$ \\
\hline
\multicolumn{7}{c}{Ultra-soft AGN spectrum ($\alpha_{\rm EUV}=-2.2$)} \\
\hline
$-3$   & $2.5$ & $2.70$ & 0.02 & $3.2_{-0.1} ^{+0.1}$ & $3.83_{-0.09}^{+0.06}$ & $0.18 _{- 0.05 }^{+ 0.07 }$ \\
$-2.5$ & $2.3$ & $2.29$ & 0.05 & $3.0_{-0.2} ^{+0.2}$ & $3.51_{-0.26}^{+0.14}$ & $0.29 _{- 0.13 }^{+ 0.25 }$ \\
$-2$   & $2.1$ & $1.92$ & 0.09 & $2.8_{-0.2} ^{+0.2}$ & $3.15_{-0.29}^{+0.18}$ & $0.51 _{- 0.21 }^{+ 0.40 }$ \\
$-1.5$ & $1.9$ & $1.57$ & 0.19 & $2.6_{-0.2} ^{+0.2}$ & $2.79_{-0.32}^{+0.22}$ & $0.93 _{- 0.37 }^{+ 0.66 }$ \\
$-1$   & $1.6$ & $1.11$ & 0.36 & $2.4_{-0.3} ^{+0.2}$ & $2.42_{-0.51}^{+0.25}$ & $1.78 _{- 0.87 }^{+ 1.15 }$ \\
$-0.5$ & $1.4$ & $0.85$ & 0.84 & $2.1_{-0.4} ^{+0.2}$ & $1.90_{-0.65}^{+0.31}$ & $2.87 _{- 1.53 }^{+ 1.58 }$ \\
$0$    & $1.0$ & $0.43$ & 1.65 & $1.8_{-0.5} ^{+0.3}$ & $1.41_{-0.67}^{+0.46}$ & $5.05 _{- 2.73 }^{+ 4.01 }$ \\
\hline
\end{tabular}}
\end{table}
\setlength{\textfloatsep}{3pt}
The observations therefore constrain an emissivity-dominant unresolved dense contribution rather than a unique halo-wide clumping factor. The inferred $C$, $\langle n \rangle$, and dense-gas fractions are moments of the assumed PDF and are especially sensitive to its poorly constrained high-density cutoff. In the ideal untruncated limit ($0<n<\infty$), the corresponding curves in Figure~\ref{figure:Clump} would be slightly steeper at large $\sigma$.

\subsection{Turbulence-driven condensation and CGM weather}

The observed \heii/\ha\ ratios require a remarkably broad density distribution, implying that the line emission is dominated by a relatively small fraction of dense gas embedded within a much more diffuse medium. While our analysis cannot uniquely determine the origin of this dense component, it provides strong observational evidence that the CGM around luminous quasars is highly inhomogeneous and dynamically multiphase. Rather than representing a population of long-lived pressure-confined clouds, the emitting gas is more naturally interpreted as continuously evolving structures produced by condensation, transport, turbulent mixing, compression, and disruption. These structures may be transient, continuously regenerated, or temporarily supported by ram pressure or magnetic fields, and therefore need not remain in strict pressure equilibrium with the surrounding hot atmosphere. 
 This picture is broadly consistent with modern hydrodynamical simulations of multiphase halos. Turbulence generated by accretion, satellite interactions, galactic outflows, weak shocks, and shear-driven mixing naturally broadens the gas density distribution and continuously creates overdense structures that cool, fragment, and mix back into the ambient medium \citep[e.g.,][]{Federrath08,McCourt18,Sparre19,Gronke20}. The inferred PDF should therefore be interpreted as the emergent statistical imprint of these nonlinear processes rather than the signature of a single physical mechanism. Likewise, the large PDF widths inferred here should not be directly translated into turbulent Mach numbers through the classical isothermal relation $\sigma_s^2=\ln(1+b^2\mathcal{M}^2)$ \citep{Federrath08}, since radiative cooling, thermal instability, compression, photoionization, and line-of-sight superposition all contribute to amplifying the density contrast.

One possible physical framework relevant for the giant nebulae studied here is the chaotic cold accretion (CCA) framework \citep{Gaspari17,Gaspari20}, in which the volume-filling hot atmosphere, once approaching $t_{\rm cool}/t_{\rm eddy}\sim1$, undergoes nonlinear condensation into a multiphase medium. In this picture, uplift, turbulent stirring, and mixing continuously generate cold condensations that subsequently feed the central SMBH, while AGN feedback disperses, reheats, or re-ionizes part of the dense phase, producing a time-dependent "CGM weather". In less virialized halos at high redshift, however, similar multiphase structures may instead be sustained by cold streams, clumpy inflows, satellite interactions, and interface compression before a fully developed hot atmosphere is established. 
Within this framework, part of the observed object-to-object variation in \heii/\ha\ may reflect different evolutionary states of the multiphase CGM, ranging from relatively diffuse and highly ionized conditions to phases characterized by a more prominent emissivity-weighted dense tail \citep{Gaspari17,Gaspari20,Barbani26a,Cammelli26a}. Additional variations in AGN luminosity history, anisotropic illumination, attenuation, cloud column density, and metallicity are also expected to modulate the observed line ratios. A direct comparison with simulations will ultimately require forward modelling of synthetic \heii\ and \ha\ observations matched to the same spatial resolution, apertures, slit geometry, and sensitivity as the data, enabling quantitative comparisons of the predicted emission, morphology, kinematics, and the underlying distributions of density, column density, and ionizing flux, while accounting for the dependence of density PDFs and clumping factors on numerical resolution and weighting scheme.

\section{Summary and Conclusions}\label{sec:sumcon}

In this work, we investigated the physical properties of the cool CGM around luminous quasars by combining model constraints and observations on the resonant \lya\ transition with the non-resonant \heii\ and \ha\ emission lines. The use of non-resonant tracers is essential to reduce the degeneracies associated with the interpretation of \lya\ emission, whose observed morphology, luminosity, and kinematics can be strongly affected by resonant scattering and radiative transfer effects.

Extending previous analytical models \citep{Cantalupo19}, we first developed a suite of \texttt{CLOUDY} photo-ionization models designed to predict the \heii/\ha\ line ratio for gas illuminated by a central quasar. The models explore a broad range of gas densities, cloud sizes and distances, quasar luminosities, and EUV spectral slopes, considering both single-density clouds and lognormal density distributions. 

We then analyzed a sample of quasars at $z_{\rm med} \simeq 2.2$ which were observed with Keck/KCWI and Keck/MOSFIRE specifically for this project. The KCWI observations were used to map extended \lya\ and \heii\ emission, while the MOSFIRE follow-up provided constraints on extended \ha\ emission along slit regions selected to overlap with the brightest \heii\ and \lya\ nebular structures. From these we obtained the results to follow:
\begin{itemize}
	\item median and 16th-84th percentile range of the distribution of individual line ratios of $\heii/\ha = 0.36^{+0.10}_{-0.10}$, $\lya/\ha = 4.8^{+1.8}_{-1.4}$, and $\heii/\lya = 0.07^{+0.02}_{-0.02}$ within projected distances of $\sim 35$ pkpc from the quasars;
	\item $\heii/\ha = 0.54^{+0.32}_{-0.21}$, $\lya/\ha = 9.1_{-3.2}^{+5.3}$, and $\heii/\lya = 0.06^{+0.03}_{-0.02}$ using a complementary stacking analysis of spectra extracted from the same selected regions;
	\item The measured $\lya/\ha$ ratios support recombination as the dominant mechanism, while collisional excitation and photon pumping appear subdominant within the MOSFIRE regions.
\end{itemize}
We extended the analysis to a sample of 26 deep VLT/MUSE observations of quasars at $z \simeq 3.1-4.5$, among which 14 systems show co-spatial extended \lya\ and \heii\ nebulae. For these systems, we measured \heii/\lya\ ratios in annuli between $\sim$10 pkpc and $\sim$35 pkpc from the quasars. Applying the same methodology to the KCWI sample, we found that:
\begin{itemize}
    \item The detected, emission-weighted \heii/\lya\ component shows no evidence for strong evolution between $z \sim 2.2$ and $z \sim 3.7$, although this does not imply an invariant underlying density PDF.  
\end{itemize}
By comparing our observational results from the stacking method to our \texttt{CLOUDY} models, we found that: 
\begin{itemize}
	\item in a single density case, our models require clouds with extremely large densities for the CGM, i.e., $n>10~\rm cm^{-3}$;
    \item assuming a log-normal density distribution, the required width decreases with increasing $n_0$, from a best-fitting $\sigma=4.2$ at $n_0=0.001~\rm cm^{-3}$ to $\sigma=2.6$ at $n_0=1~\rm cm^{-3}$. Low-density models require very broad PDFs, whereas the $n_0 = 1 ~\rm cm^{-3}$ limiting case reduces the required width but shifts the tension to a globally high mean density;
    \item a softer ($\alpha_{EUV}=-2.2$) ionizing AGN spectrum reduces the density dispersion required to reproduce the observed \heii/\ha\ ratio.
\end{itemize}
Our results indicate that the cool quasar CGM is multiphase and that its recombination emission is dominated by unresolved dense gas. Rather than reflecting a unique static halo-wide density PDF, the inferred density distribution likely represents an emissivity-weighted snapshot of a time-dependent CGM weather cycle driven by cooling, turbulence, and mixing. This picture is qualitatively consistent with CCA once a turbulent hot atmosphere is established, while cold streams and interaction-driven compression may dominate in less virialized high-redshift halos. Quantitative tests will require synthetic \heii\ and \ha\ observables matched to the spatial resolution, apertures, and sensitivity of the data.

\begin{acknowledgements}
This project was supported by the European Research Council (ERC) Consolidator Grant 864361 (CosmicWeb) and by INAF PRIN 2022 2022TKPB2P "BIG-z". M.G. acknowledges support from the ERC Consolidator Grant BlackHoleWeather (101086804). CF acknowledges support from the Ricerca Fondamentale INAF 2023 Data Analysis grant 1.05.23.03.04 "ARCHIE ARchive Cosmic HI \& ISM  Evolution".
\end{acknowledgements}
\bibliographystyle{aa} 
\small{
\linespread{0.95}\selectfont
\bibliography{andrea.bib}
}

\appendix

\section{Temperature - Density}\label{sec:temp}

\begin{figure}[h]
   \begin{center}
   \includegraphics[height=0.45\textheight,angle=0]{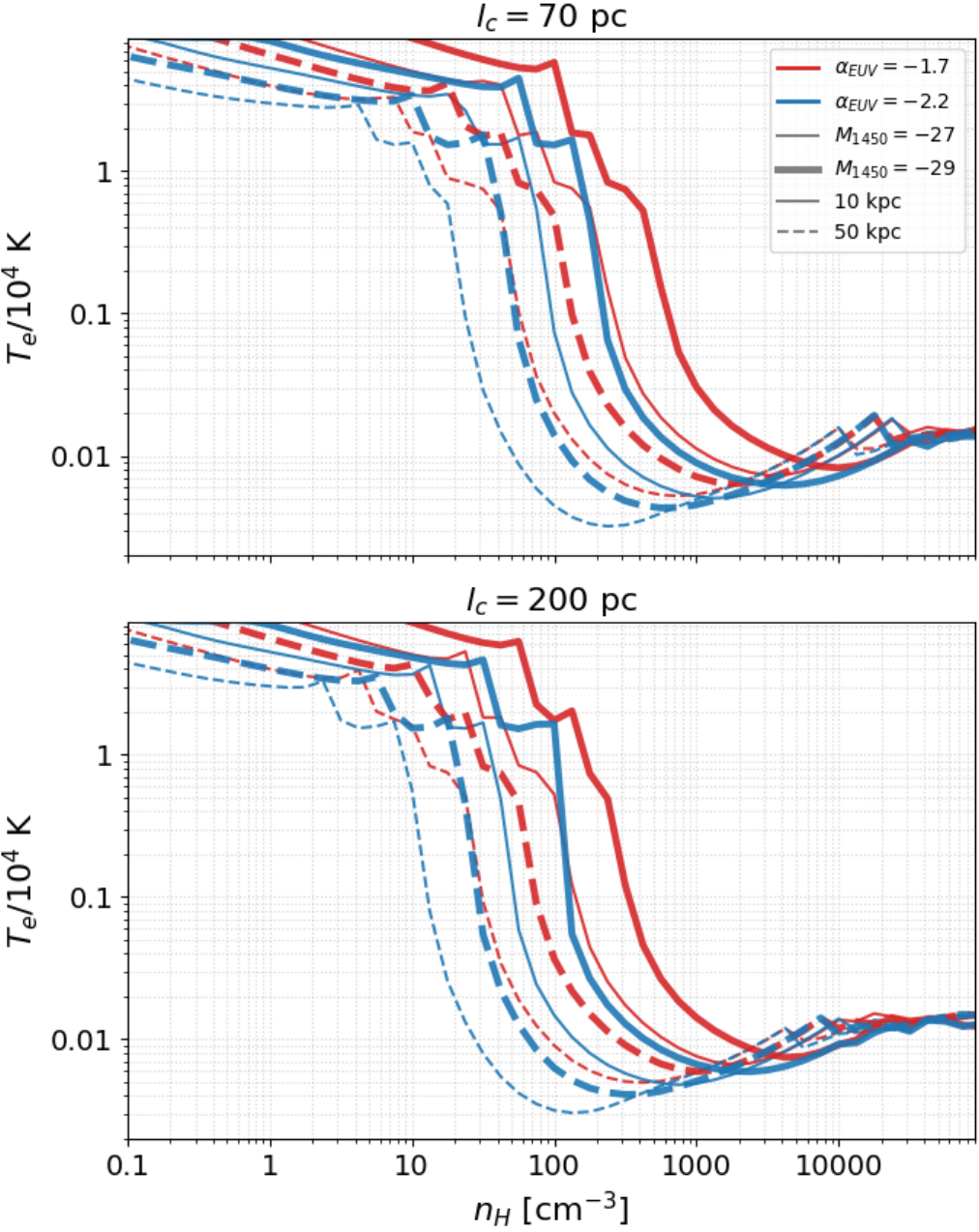}
   \caption{Electron temperature as a function of hydrogen density for the same photo-ionization models shown in Figure~\ref{figure:constDen}. Colors indicate the EUV spectral slope ($\alpha_{\rm EUV}=-1.7$ in red and $-2.2$ in blue), while line thickness and style distinguish AGN luminosities at 1450 \AA, cloud size, and distances from the ionizing source. The temperature evolution reflects the progressive confinement of the hot \heiii\ region as the clouds become optically thick.}
   \label{fig:Tden}
   \end{center}
\end{figure}

Figure~\ref{fig:Tden} provides additional insight into the thermal structure of the emitting gas by showing the electron temperature ($T_e$) as a function of density ($n_e$) for the same set of models. 
The temperature evolution closely follows the changes observed in the \heii/\ha\ ratio, confirming that the transition from the optically thin to the optically thick regime is associated with a progressive confinement of the \heiii-emitting gas to the innermost regions of the clouds, while the outer layers become cooler and dominated by \hii\ emission.
At low densities, where the gas is nearly fully ionized, $T_e$ remains relatively high ($T_e > 2\times10^4$ K), particularly for harder ionizing spectra, reflecting efficient photoheating by high-energy photons. As the density increases and the \heiii\ region becomes progressively confined to the inner cloud layers, $T_e$ rapidly decreases because the outer regions are predominantly heated by lower-energy photons and become dominated by \hii\ emission. 
At the highest densities, all models converge toward $T_e \approx 10^2$ K, indicating that cooling becomes highly efficient once the clouds are deeply optically thick ($n_e>5\times 10^3~\rm cm^{-3}$). In this regime, the strong dependence of the \heii/\ha\ ratio on $\alpha_{\rm EUV}$ (see Figure~\ref{figure:constDen}) is therefore primarily driven by differences in the penetration depth of photons above 54.4 eV, which regulate the extent of the \heiii\ region, rather than by large variations in the gas temperature.

\setlength{\textfloatsep}{-2pt}

\section{Finite-Grid Clumping Factor}\label{appendix:finitegrid}

The standard relation $C=\exp(\sigma^2)$ assumes that the log-normal density distribution extends over the full interval $0<n<\infty$. In our CLOUDY models, however, the density distribution is sampled only within the finite range $10^{-5}~\rm cm^{-3}\leq n\leq 10^5~\rm cm^{-3}$. We therefore derive the clumping factor associated with the corresponding truncated log-normal distribution. 
Introducing the variable $x \equiv \ln(n)$ the untruncated probability density function is
\begin{equation}
PDF(x)= \frac{1}{\sqrt{2\pi}\sigma} \exp\left[-\frac{(x-\mu)^2}{2\sigma^2} \right],
\end{equation}
where $\mu=\ln(n_0)$ and $n_0$ is the median density. For a finite density interval between $x_{\min}=\ln(n_{\rm min})$ and $x_{\rm max}=\ln(n_{\rm max})$, the normalization of the truncated distribution is
\begin{equation}
Z = \int_{x_{\min}}^{x_{\max}}PDF(x) \,dx.
\end{equation}
Defining
\begin{equation}
\alpha = \frac{\ln(n_{\rm min})-\mu}{\sigma},
\qquad
\beta = \frac{\ln(n_{\rm max})-\mu}{\sigma},
\end{equation}
where $\Phi$ is the cumulative distribution function of a standard normal distribution:
\begin{equation}
\Phi(y) \equiv \frac{1}{\sqrt{2\pi}} \int_{-\infty}^{y} \exp\left(-\frac{t^2}{2}\right)\,dt,
\end{equation} 
and the normalization can be written as
\begin{equation}
Z=\Phi(\beta)-\Phi(\alpha).
\end{equation}
Therefore, the truncated probability density function is
\begin{equation}
PDF_{\rm tr}(x)=\frac{PDF(x)}{Z}.
\end{equation}
The $k$-th moment of the density distribution is
\begin{equation}
\langle n^k\rangle_{\rm tr} = \frac{1}{Z} \int_{x_{\rm min}}^{x_{\rm max}} e^{kx}p(x) \, dx.
\end{equation}
Evaluating the truncated moments for $k=1$ and $k=2$, and using $C_{\rm tr} \equiv \langle n^2\rangle_{\rm tr} / \langle n\rangle_{\rm tr}^2$, we obtain
\begin{equation}
    C_{\rm tr} = e^{\sigma^2} \frac{Z_0 Z_2}{Z_1^2}.
\end{equation}
In the limit $n_{\rm min}\rightarrow 0$ and $n_{\max}\rightarrow\infty$, one has $Z_0=Z_1=Z_2=1$, and the standard expressions $\langle n\rangle=n_0e^{\sigma^2/2}$ and $C=e^{\sigma^2}$ are recovered. For finite density limits, however, this asymptotic behavior no longer holds. As $\sigma$ increases, the truncated $C_{\rm tr}$ initially rises because progressively more probability is shifted toward high densities. Once the PDF becomes sufficiently broad, an increasing fraction of the high-density tail lies beyond $n_{\rm max}$ and is excluded from the integration. Since $\langle n^2\rangle$ is dominated by the densest gas, $C_{\rm tr}$ reaches a maximum and can subsequently decrease despite the increasing PDF width.

\section{Skewness impact on the log-normal distribution}\label{appendix:skewness}

The skewed log-normal PDF is described by the formula to follow
\begin{equation} \label{eqPDF}
	PDF(\ln(n)) ~d\ln(n) = \frac{\Phi(\alpha_{\rm sk}, \ln(n), \sigma)}{\sqrt{2 \pi} \sigma}  e^{-\frac{(\ln(n) - \mu)^2}{2 \sigma^2}}  ~d \ln(n),
\end{equation}
where
\begin{equation}
	\Phi(\alpha_{\rm sk}, \ln(n), \sigma) = [ 1 + \rm{erf} ( \alpha_{\rm sk} (\ln(n) - \mu) / \sqrt{2}\sigma) ],
\end{equation}
with $\alpha_{\rm sk}$ the skewness parameter. 
While $\alpha_{\rm sk} = 0$ sets a classical log-normal density distribution, in the case $\alpha_{\rm sk} \neq 0$, the average density can be derived by evaluating the integral $\langle n \rangle_c = \int_0^{\infty} n \, PDF(n)\, dn$, where $PDF(n)$ is the skewed log-normal distribution. Using the integral identity reported in Eq.~(8.259.1) of \citet{Gradshteyn07}, we obtain
\begin{equation}
\langle n \rangle = n_0 e^{\sigma^2/2} \, f(\sigma, \alpha_{\rm sk}),
\end{equation}
where $f(\sigma, \alpha_{\rm sk})$ is a correction factor that accounts for the skewness of the distribution, given by
\begin{equation}
f(\sigma,\alpha_{\rm sk}) = 1+\rm{erf} [ \alpha_{\rm sk}\sigma / \sqrt{2(1+\alpha_{\rm sk}^2)}].
\end{equation}
In general, $\alpha_{\rm sk} > 0$ (<0) enhances the contribution of high(low)-density regions, increasing (decreasing) $\langle n \rangle _c$ at fixed $n_0$. Since the clumping factor is defined as $C = \langle n^2 \rangle_c / \langle n \rangle_c^2$, it can be expressed as a function of $\sigma$ and $\alpha_{\rm sk}$ as
\begin{equation}\label{eqgsigalpha}
C = e^{\sigma^2} \, g(\sigma, \alpha_{\rm sk}),
\end{equation}
where $g(\sigma, \alpha_{\rm sk}) = f(2\sigma,\alpha_{\rm sk})/f^2(\sigma,\alpha_{\rm sk})$.\\

\begin{figure}[t]
   \begin{center}
   \includegraphics[height=0.29\textheight,angle=0]{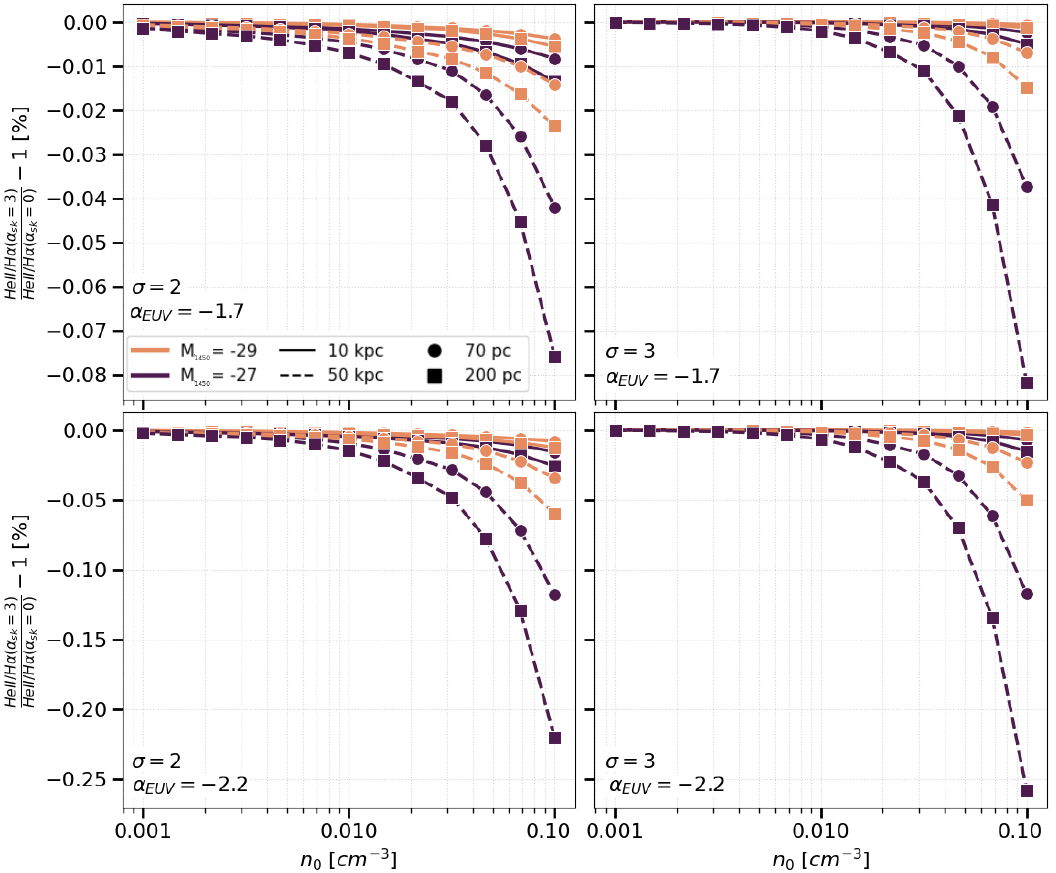} 
   \caption{Relative variation (in percent) of the \heii/\ha\ emission-line ratio induced by the introduction of skewness in the density distribution by a factor $\alpha_{\rm sk}=3$, which represents the maximum skewness considered. The results are shown as a function of the characteristic density $n_0$ for different values of $\sigma$, $\alpha_{\rm EUV}$, $M_{1450}$, cloud distance $D_{\rm c}$, and cloud size $l_c$, as indicated.}
   \label{figure:lognormskewn}
   \end{center}
\end{figure}

To evaluate the impact of $\alpha_{\rm sk}$ on lowering the \heii/\ha\ line ratio, we present, in Figure~\ref{figure:lognormskewn}, the relative variation (in percentage) of the \heii/\ha\ ratio. This is computed as the ratio obtained for $\alpha_{\rm sk}=3$, corresponding to the maximum asymmetry estimated in \Laz, divided by that obtained for $\alpha_{\rm sk}=0$, minus one. We tested these differences for two specific $\sigma=2$ and 3. We find that the \heii/\ha\ line ratio decreases by up to $\sim 0.2 - 0.25\%$ for the parameter combination $\alpha_{\rm EUV} = -2.2$, $M_{1450} = -27$, $D_c = 50\,\mathrm{kpc}$, and $l_c = 200\,\mathrm{pc}$, listed here in decreasing order of impact on the line ratio.
We make publicly available a repository of tables reporting all the \heii/\ha\ line ratios as a function of the model parameters explored in this work. The grid spans the same parameter space adopted throughout the paper, including $\alpha_{\rm EUV}$, $M_{1450}$, cloud distance $D_c$, cloud size $l_c$, and characteristic gas density $n_0$. 

\setlength{\textfloatsep}{12pt}

To facilitate comparison between our photo-ionization models and observations, we make publicly available a repository containing the full grid of \heii/\ha\ line ratios explored in this work. This includes a sampling of the skewness parameter over $\alpha_{\rm sk} = 0-3$ (in steps of 0.25) and the dispersion over $\sigma = 0-4$ (in steps of 0.05). For each parameter combination, we obtain the line ratio by integrating over the skewed log-normal density distribution. The full table is provided in electronic form\footnote{\url{https://github.com/AndTrx/CGM-multiTransitions}}.

\section{KCWI and MOSFIRE data} \label{app:data}

The top panel in Table~\ref{tab:KCWI} lists the quasars observed with Keck/KCWI, their coordinates, apparent magnitude in the $i$ band, and redshift, while in the bottom panel we report the main observational properties of KCWI data, including the number of exposures, total exposure time, and seeing (FWHM).

\begin{table}
\caption{Observational logs for the quasar sample at $z \approx 2.2$. The top panel lists target coordinates, $i$-band magnitudes, and redshifts. The bottom panel shows the dates, exposure times, and seeing conditions (FWHM) of the Keck/KCWI observations.}
\label{tab:KCWI}
\centering

\begin{adjustbox}{width=9cm,center}
\begin{tabular}{l c c c c}
\hline 
\vspace{-0.3cm}\\
\multicolumn{5}{c}{Sample quasars at $z \approx 2.2$} \\
\vspace{-0.3cm}\\
\hline 
\vspace{-0.3cm}\\
Target & RA & Dec & $mag~i$ & z  \\
       & [HMS] & [DMS] & &  \\
\hline
CH-Q01   & 02:09:45.60 & +00:53:04.00 & 18.2  & 2.190  \\
CH-Q02   & 02:09:57.67 & +00:51:00.00 & 19.58 & 2.176  \\
J0010+06 & 00:10:05.75 & +06:17:20.64 & 18.50 & 2.263 \\
J0030+05 & 00:30:21.81 & +05:30:52.94 & 18.14 & 2.270  \\
J0141+07 & 01:41:28.26 & +07:06:06.13 & 18.33 & 2.265  \\
J0338-06 & 03:38:58.47 & -06:08:32.68 & 17.97 & 2.266  \\
J2316+09 & 23:16:49.55 & +09:06:34.74 & 18.01 & 2.264  \\
Q0050+00 & 00:50:21.22 & +00:51:35.00 & 17.82 & 2.247  \\
Q1130+11 & 11:30:14.51 & +11:48:59.15 & 18.01 & 2.264  \\
Q1227+28 & 12:27:27.47 & +28:48:47.88 & 17.73 & 2.266  \\
Q1230+33 & 12:30:35.46 & +33:20:00.48 & 17.76 & 2.315  \\
Q1411+00 & 14:11:23.51 & +00:42:53.01 & 17.96 & 2.269  \\
Q1426+25 & 14:26:35.86 & +25:55:23.69 & 17.59 & 2.255  \\
\hline
\vspace{-0.3cm}\\
\multicolumn{5}{c}{Keck/KCWI} \\
\hline 
Target & Date & $N_{\rm exp}$ & $t_{\rm exp}$ & FWHM \\
  & & & [h] & [arcsec] \\
\hline   
CH-Q01   & 2025-10-13 & 4 & 1.0 & 1.33--1.68 \\
CH-Q02   & 2025-10-13 & 4 & 1.0 & 1.27--1.61 \\
J0010+06 & 2025-08-29 & 8 & 2.2 & 1.22--1.60 \\
J0030+05 & 2025-10-13/14 & 9 & 2.2 & 1.54--1.84 \\
J0141+07 & 2025-10-14 & 4 & 1.0 & 1.33--1.52 \\
J0338-06 & 2025-10-13/14 & 5 & 1.1 & 1.38--1.42 \\
J2316+09 & 2025-10-13/14 & 8 & 2.0 & 1.41--2.08 \\
Q0050+00 & 2025-08-29 & 7 & 1.9 & 1.04--1.29 \\
Q1130+11 & 2025-03-25 & 3 & 0.8 & 1.34--1.40 \\
Q1227+28 & 2025-03-24 & 7 & 1.9 & 1.19--1.36 \\
Q1230+33 & 2025-03-24/25 & 8 & 2.2 & 1.79--2.80 \\
Q1411+00 & 2025-03-25 & 3 & 0.83 & 2.1--2.13 \\
Q1426+25 & 2025-03-24/25 & 8 & 2.2 & 1.11--1.61 \\
\hline
\end{tabular}
\end{adjustbox}
\end{table}

\begin{figure*}[t]
   \begin{center}
   \includegraphics[height=0.6\textheight,angle=0]{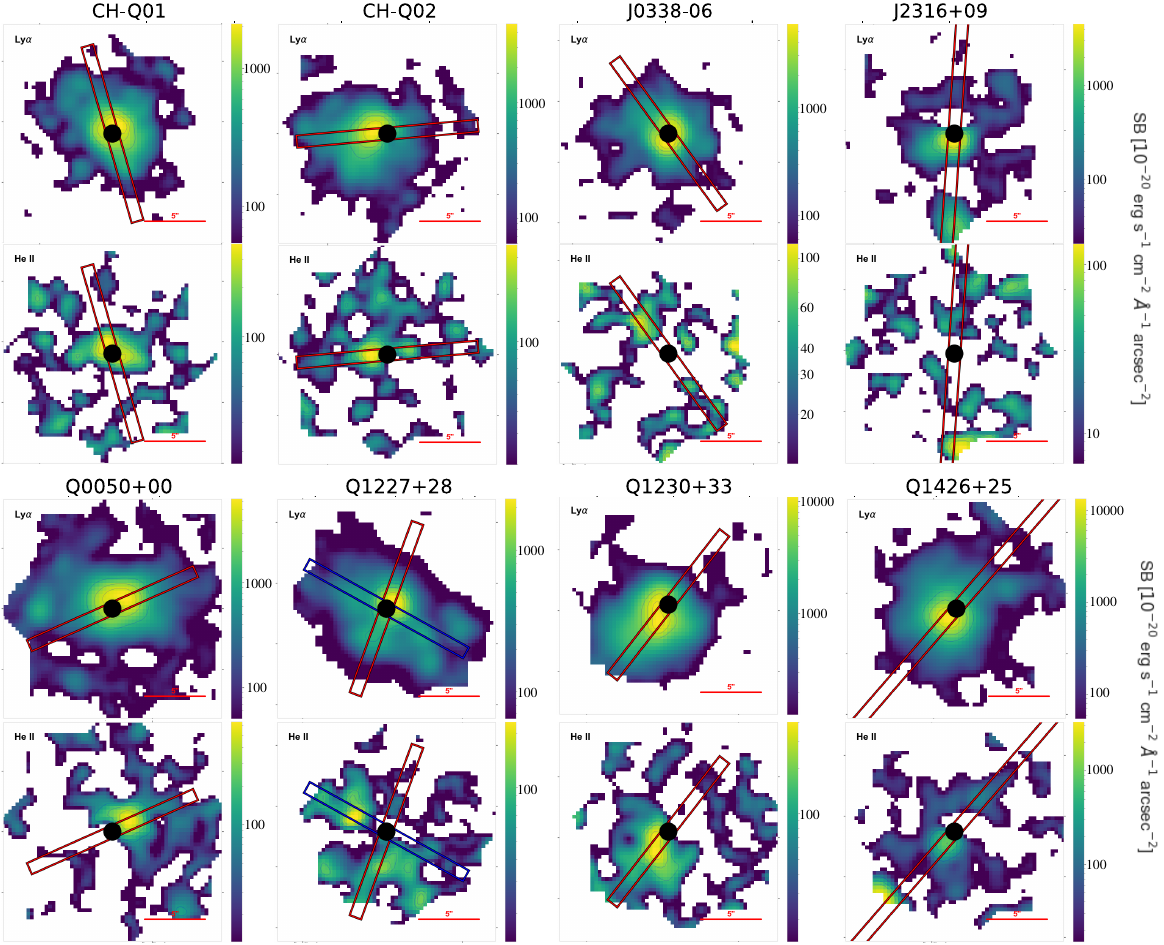} 
    \caption{SB maps (zeroth moment of the flux distribution) of the extended \lya\ and \heii\ nebulae detected around the $z_{\rm med} \simeq 2.2$ quasars observed with Keck/KCWI. The overlaid rectangles indicate the corresponding MOSFIRE slit configurations adopted for the follow-up \ha\ observations. The length of the red line is 5$\arcsec$, corresponding to $\approx 40~\rm pkpc$.}
\label{appfig:slits}
   \end{center}
\end{figure*}

In Section~\ref{sec:anaKeck}, we describe the procedure adopted to extract the extended \lya\ and \heii\ nebulae from the sample of 13 quasars observed with Keck/KCWI. Based on the morphology and SB distribution of the detected emission, we selected the systems showing the most significant extended \lya\ and \heii\ emission to design the MOSFIRE slit configurations. The slits were positioned and oriented to maximize the overlap with the brightest regions of the nebulae, to maximize the probability of detecting extended \ha\ emission. Figure~\ref{appfig:slits} shows the KCWI SB maps of the \lya\ and \heii\ emission together with the corresponding MOSFIRE slit positions for each target. Table~\ref{tab:MOSFIRE} summarizes the MOSFIRE observing log for the $z \approx 2$ quasar sample.

\begin{table}
\caption{Observational logs for the quasar sample at $z \approx 2.2$. Summary of the Keck/MOSFIRE observations, including dates, exposure times, and seeing conditions (FWHM).}
\label{tab:MOSFIRE}
\centering
\begin{adjustbox}{width=8cm,center}
\begin{tabular}{lcccc}
\hline
\multicolumn{4}{c}{Keck/MOSFIRE} \\
\hline 
Target & Night & $t_{\rm exp}$ & FWHM \\
  & & [h] & [arcsec]  \\
\hline
CH-QSO1  & 2025-12-27       & 1.7  & 0.88 \\
CH-QSO2  & 2025-12-25       & 2.0  & 0.62 \\
J0338-06 & 2025-12-25, 26   & 2.2  & 0.74 \\
J2316+09 & 2025-12-25, 26   & 2.0  & 0.62 \\
Q0050+00 & 2025-12-24, 26   & 2.0  & 1.05 \\
Q1227+28(S1) & 2025-06-03,04 & 1.8 & 0.82 \\
Q1227+28(S2a) & 2025-06-03 & 1.0\textsuperscript{a} & 1.20 \\
Q1230+33 & 2025-06-04 & 2.2 & 0.74 \\ 
Q1426+25 & 2025-06-03 & 0.3\textsuperscript{b} & 0.97 \\
SLUG+Rc  & 2025-12-27 & 2.5  & 0.51 \\
\hline
\multicolumn{4}{c}{Archival Keck/MOSFIRE} \\
\hline
Q1227+28 & 2019-05-22 & 1.5 & 0.7 \\ 
J0010+06     & 2019-11-04 & 1.5 & 0.7 \\ 
SLUG      & 2014-10-02 & 4.8 & 0.7 \\ 
\hline
\end{tabular}
\end{adjustbox}
\footnotesize
\textsuperscript{a} Sequence affected by thick clouds toward the end of observations.\\
\textsuperscript{b} Limited integration (21 min); data affected by poor cosmic-ray rejection and persistence.
\end{table}

\begin{figure*}[t]
   \begin{center}
   \includegraphics[height=0.42\textheight,angle=0]{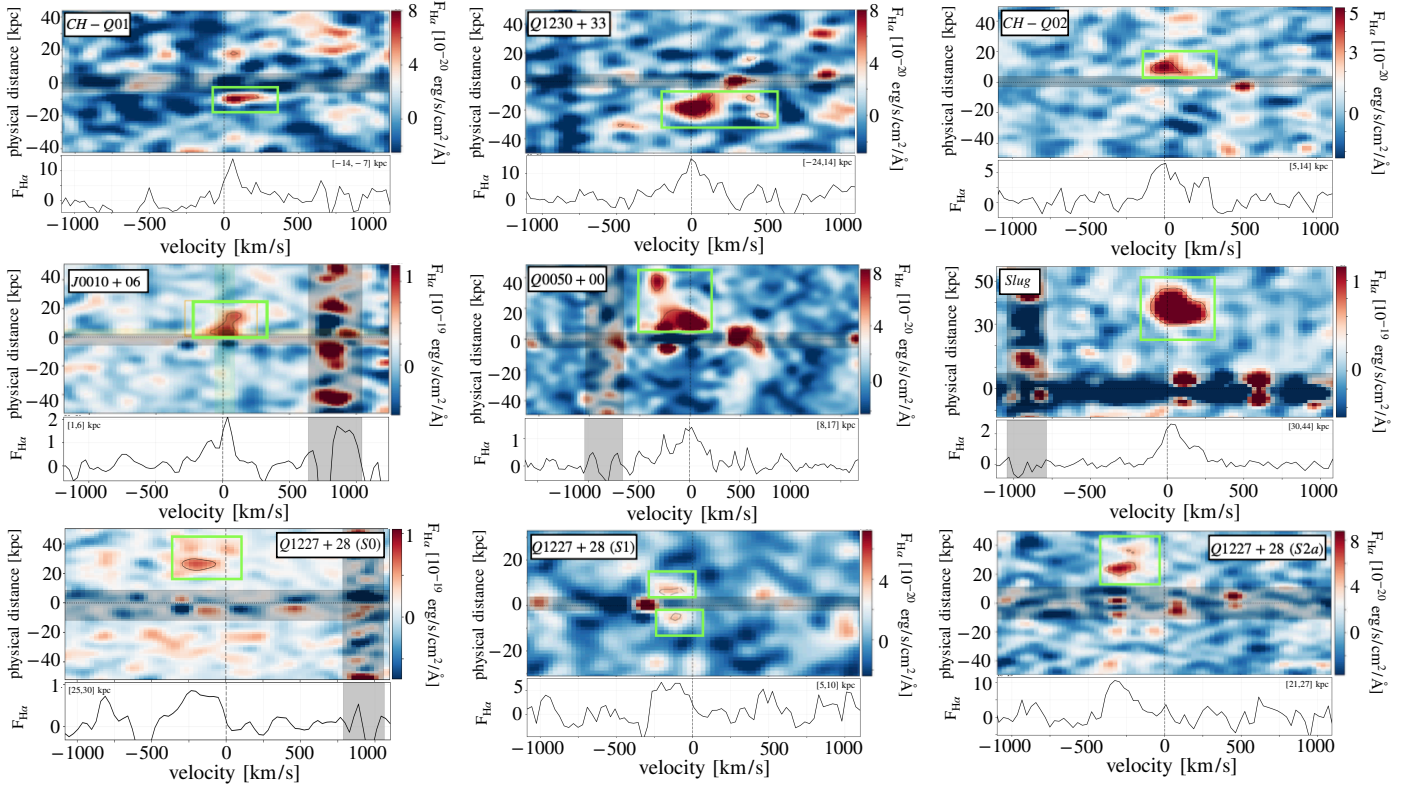} 
   \caption{Two-dimensional MOSFIRE spectra of the \ha\ emission line in cases of successful detection. The primary emission features are enclosed in green boxes. The sub-panels display the extracted 1D flux profiles. The right panels show the flux profile at the spectral region of the \ha\ detection. The semi-transparent shaded areas identify regions with significant noise or artifacts, specifically those originating from PSF subtraction or sky-line residuals.}
   \label{appfigure:CGMHaLyaHe}
   \end{center}
\end{figure*}

For each target, the Keck/KCWI integral-field data were converted into synthetic slit spectra, matching the position and orientation of the corresponding Keck/MOSFIRE observations (see Section~\ref{sec:anaKeck} for details). Figure~\ref{appfigure:CGMHaLyaHe} presents the two-dimensional MOSFIRE slit spectra for the objects where extended \ha\ emission is detected, as highlighted by the green boxes. Each panel is complemented by two one-dimensional extractions: a lower sub-plot showing the 1D spectrum (flux versus velocity shift) and a right-hand sub-plot displaying the spatial flux profile (flux versus physical offset from the quasar). The specific spatial and spectral integration intervals used for these extractions are explicitly labeled above the respective sub-plots. The semi-transparent black regions mask spectral and spatial intervals potentially affected by sky-line residuals or artifacts from the PSF subtraction process.

The extended \ha\ emission is typically detected at physical offsets ranging from 5 to 50 kpc from the central quasar. An exception is found in the field of J0338-06, where a potential \ha\ source is identified at a much larger distance of approximately 150 kpc. However, this feature is consistent with a compact \ha\ line-emitter rather than part of the diffuse gaseous nebula; therefore, it is excluded from our current estimates of the CGM emission. 
Table~\ref{tab:final_line_ratios} reports the line fluxes and ratios measured for the targets with both Keck/KCWI and MOSFIRE observations, while the corresponding ratios are shown in Figure~\ref{figure:HeHaLya1}. The individual measurements display substantial object-to-object variation among the targets with reliable detections. The \lya/\ha\ ratio ranges from approximately 2.3 to 12.5, while \heii/\ha\ spans approximately 0.2-1.3. The three slit configurations available for Q1227+28 also reveal significant spatial variations, particularly in \lya/\ha\ and \heii/\ha, illustrating that the measured ratios depend on the specific regions sampled by the MOSFIRE slit. The ratios measured from the stacked spectra are broadly consistent with these values, although the larger stacked \lya/\ha\ ratio reflects the different weighting of individual systems in a flux-weighted stack.

\begin{table}
\centering
\caption{Line fluxes and line ratios measured within the area of the MOSFIRE slits and within a distance of 35 kpc from the quasars for our sample at $z\sim2.3$. Fluxes are reported in units of $10^{-17}\,{\rm erg\,s^{-1}\,cm^{-2}}$. For Q1227+28, we report both the combined value, obtained by averaging the three available MOSFIRE slit measurements, and the individual slit measurements. The row "All targets with MC method" reports the median ratios and 16th-84th percentile ranges estimated through the Monte Carlo procedure described in the text. The final row reports the line fluxes and ratios measured from the stacked one-dimensional spectra. The column "Stack" indicates whether a given target is included in the stacked-spectrum analysis.}
\label{tab:final_line_ratios}
\scriptsize
\setlength{\tabcolsep}{2.2pt}
\renewcommand{\arraystretch}{1.15}
\resizebox{\columnwidth}{!}{
\begin{tabular}{lccccccc}
\hline
Name & $F_{\lya}$ & $F_{\ha}$ & $F_{\heii}$ & $\lya/\ha$ & $\heii/\ha$ & $\heii/\lya$ & Stack \\
\hline 
CH-Q01 & $5.47 \pm 0.36$ & $1.67 \pm 0.20$ & $0.63 \pm 0.09$ & $3.28 \pm 0.46$ & $0.38 \pm 0.07$ & $0.115 \pm 0.019$ & Y \\ 
CH-Q02 & $11.79 \pm 0.33$ & $0.95 \pm 0.13$ & $1.19 \pm 0.09$ & $12.45 \pm 1.73$ & $1.25 \pm 0.20$ & $0.100 \pm 0.008$ & Y \\ 
J0338-06 & $4.42 \pm 0.71$ & $<$0.31 & $0.18 \pm 0.09$ & $>14.11$ & $>$0.5 & $0.04 \pm 0.02$ & Y \\ 
Q0050+00 & $16.12 \pm 1.12$ & $5.83 \pm 0.52$ & $1.19 \pm 0.18$ & $2.77 \pm 0.31$ & $0.21 \pm 0.04$ & $0.074 \pm 0.012$ & Y \\ 
Q1230+33 & $28.01 \pm 1.69$ & $2.38 \pm 0.34$ & $0.92 \pm 0.08$ & $11.79 \pm 1.81$ & $0.39 \pm 0.06$ & $0.033 \pm 0.003$ & Y \\ Q1426+25$^{a}$ & $14.90 \pm 1.86$ & -- & $1.21 \pm 0.13$ & -- & -- & $0.081 \pm 0.013$ & N \\ 
J0010+06 & $7.77 \pm 0.78$ & $3.32 \pm 0.37$ & $1.40 \pm 0.22$ & $2.34 \pm 0.35$ & $0.42 \pm 0.08$ & $0.180 \pm 0.034$ & Y \\ 
SLUG-Rc$^{b}$ & $12.50 \pm 1.12$ & $3.84 \pm 0.16$ & $0.63 \pm 0.03$ & $3.26 \pm 0.32$ & $0.16 \pm 0.01$ & $0.050 \pm 0.005$ & N \\
SLUG$^{b}$ & $14.41 \pm 1.73$ & $2.62 \pm 0.21$ & -- & $5.50 \pm 0.79$ & -- & -- & N \\
Q1227+28$^{c}$ & $10.79 \pm 0.96$ & $2.16 \pm 0.12$ & $0.53 \pm 0.03$ & $5.00 \pm 0.53$ & $0.25 \pm 0.02$ & $0.049 \pm 0.005$ & Y \\
Q1227+28-S0$^*$ & $9.64 \pm 1.24$ & $3.39 \pm 0.19$ & $0.67 \pm 0.09$ & $2.84 \pm 0.40$ & $0.20 \pm 0.03$ & $0.069 \pm 0.013$ & Y \\ 
Q1227+28-S1$^*$ & $11.91 \pm 0.79$ & $0.90 \pm 0.15$ & $0.54 \pm 0.05$ & $13.17 \pm 2.37$ & $0.60 \pm 0.12$ & $0.045 \pm 0.005$ & Y \\ 
Q1227+28-S2a$^*$ & $11.30 \pm 1.29$ & $2.30 \pm 0.22$ & $0.47 \pm 0.03$ & $4.92 \pm 0.73$ & $0.20 \pm 0.02$ & $0.042 \pm 0.006$ & Y \\
\hline
\multicolumn{8}{c}{Median and 16th/84th percentiles} \\
\hline
\multicolumn{4}{l}{All targets with MC method (Figure~\ref{figure:HeHaLya1})}  & $4.8 _{-1.4}^{+1.8}$ & $0.36 _{-0.1}^{+0.1}$ & $0.07 _{-0.02}^{+0.02}$ & -- \\
\hline
\hline
\multicolumn{8}{c}{Stacked spectrum} \\
\hline
Stack & $15.80^{+4.03}_{-3.60}$ & $1.80^{+0.60}_{-0.68}$ & $0.94^{+0.26}_{-0.26}$ & $9.05_{-3.24}^{+5.30}$ & $0.54_{-0.21}^{+0.32}$ & $0.060_{-0.020}^{+0.025}$ &  \\
\hline
\end{tabular}
}
\begin{flushleft}
\scriptsize
$^{a}$ This quasar does not have a reliable \ha\ detection because of the limited MOSFIRE exposure time.
$^{b}$ For the SLUG and SLUG-Rc, the line fluxes are taken from the literature \citep{Leibler18,Cantalupo19}; these objects are not part of our main sample and are not included in the stack.
$^{c}$ For Q1227+28 we report the average values obtained from the three available MOSFIRE slit measurements.
$^*$ Estimates from three different slits of the same target. Figure~\ref{figure:HeHaLya1} reports the average between these three values.\\
For the main sample, fluxes are averaged over a series of independent extractions; uncertainties combine the propagated measurement errors and the scatter among the extractions. The stacked spectrum is continuum-zero subtracted before flux extraction.
\end{flushleft}
\end{table}

\section{Monte Carlo Framework for Robust Line Ratio Estimation}\label{app:MC}


In this Appendix, we describe the fully automated Monte Carlo (MC) framework, which is designed to estimate line fluxes and their uncertainties and to explore the full range of plausible extraction configurations while preserving physical consistency across all transitions. Starting from three aligned slit spectra (\ha\ from MOSFIRE and \lya\, \heii\ from IFS), we define a common spatial extraction region extending to $\pm 33$ kpc from the quasar position, excluding a central PSF-dominated region.
For each source, we generate $10000$ MC realizations in which we simultaneously vary the key extraction parameters within physically motivated priors. These include: (i) the size of the central PSF mask, uniformly sampled between $2$ and $10$ kpc; (ii) the position of a single background window, whose center is allowed to vary between $-60$ and $+60$ kpc along the slit; (iii) the width of this background region, sampled between $5$ and $20$ kpc; and (iv) independent velocity integration windows for each emission line. The velocity windows are defined by a central velocity and a half-width, with priors of $-300$ to $+300$ km s$^{-1}$ and $100-800$ km s$^{-1}$ for \ha\ and \heii, and $-600$ to $+600$ km s$^{-1}$ and $200-1600$ km s$^{-1}$ for \lya. To ensure a clean separation between signal and background, we impose a minimum spatial offset of $2$ kpc between the background region and the flux extraction region. 
For each MC realization, the line flux is computed by integrating the signal within the selected spatial and spectral window, after subtracting a local background estimated from the same velocity interval but within the displaced background region. Before this procedure, the data are optionally cleaned of cosmic-ray residuals using a conservative sigma-clipping filter ($7\sigma$ for \ha\ and \heii, $8\sigma$ for \lya), ensuring that only extreme, isolated outliers are removed without affecting extended emission.
This approach produces, for each source, joint distributions of \ha, \lya, and \heii\ fluxes corresponding to consistent sets of extraction parameters. From these, we derive the distributions of the line ratios \lya/\ha, \heii/\ha, and \heii/\lya. The final reported values are obtained from the subset of realizations where all ratios lie within a physically plausible range ($>10^{-3}$). We define a representative best solution as the realization closest to the median of the three ratios in logarithmic space, ensuring that all reported parameters and fluxes correspond to a single, self-consistent configuration. 
This methodology allows us to robustly propagate systematic uncertainties associated with extraction choices, while avoiding biases introduced by subjective parameter tuning, and provides a statistically consistent framework for comparing line ratios across the sample.

\section{\heii\ emission lines in nebulae and statistics of the detections} \label{app:spec}

For each object, we extracted circularly averaged \lya\ and \heii\ surface-brightness profiles around the adopted quasar position from the optimally extracted narrow-band images. We converted radii to comoving kpc using the source redshift, and corrected the profiles for cosmological surface-brightness dimming by multiplying by $(1+z)^4$. Individual profiles were then interpolated onto a common radial grid and combined using the median, with the 25th and 75th percentiles used to represent the object-to-object scatter. The \heii/\lya\ ratio was computed for each object before stacking the individual ratio profiles.
\begin{figure}[t]
   \begin{center}
   \includegraphics[height=0.4\textheight,angle=0]{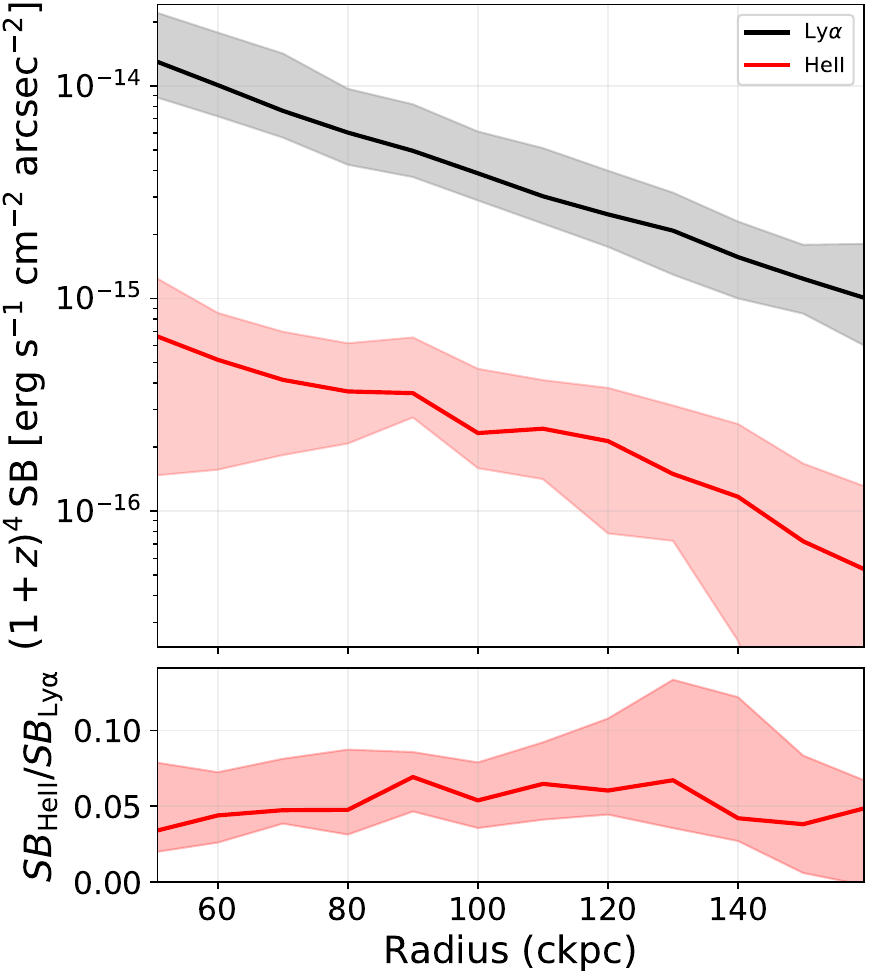} 
   \caption{Median radial surface-brightness profiles of \lya\ and \heii\ for the combined KCWI+MUSE sample. The upper panel shows the cosmological-dimming-corrected profiles, $(1+z)^4{\rm SB}$, while the lower panel shows the corresponding radial \heii/\lya\ surface-brightness ratio. Shaded regions indicate the 25th-75th percentile range among the individual objects.}
   \label{figure:RatioHeIILyaSB}
   \end{center}
\end{figure}
We therefore assume that the regions producing \heii\ emission are approximately co-spatial with those producing the hydrogen emission. This assumption is supported by the very similar radial surface-brightness profiles of the two lines, which mainly differ by an approximately constant normalization factor, with ${\rm SB}_{\rm HeII}/{\rm SB}_{\rm Ly\alpha}\simeq0.05$ over the probed radial range (Figure~\ref{figure:RatioHeIILyaSB}).\\

\begin{table}
\caption{
Properties of the quasar sample observed with MUSE ($\sim 4~\rm hr$ per target). $F_{\rm rms}$ indicates the $1\sigma$ SB limit measured at 5500~\AA\ within a 0.7~arcsec aperture.}
\label{tabapp:magg}
\bigskip
\centering\small\setlength\tabcolsep{3pt}
\begin{adjustbox}{width=8.6cm,center}
\begin{tabular}{cccccc}
\hline
\vspace{-0.3cm}\\
\multicolumn{5}{c}{Sample quasars at $z_{\rm med} \approx 3.671$} \\
\hline
Names & RA & DEC & $z_{\rm{QSO}}$\textsuperscript{a} & $mag_r$\textsuperscript{b} & $F_{rms}$ \\ 
      & (hh:mm:ss) & (dd:mm:ss) &    &   & $10^{-20}~ \rm cgs$   \\ 
\hline
J0106          &   01:06:19.24 &   +00:48:23.31   &   4.4402 &   19.10 &  1.84  \\                          
J0124          &   01:24:03.77 &   +00:44:32.76   &   3.8359 &   17.95 &  1.99 \\                            
J0133          &   01:33:40.31 &   +04:00:59.77   &   4.1709 &   18.45 &  1.86   \\                          
J0137          &   01:37:24.36 &   -42:24:17.30   &   3.975   &   18.46 &  1.41   \\                         
J0157          &   01:57:41.56 &   -01:06:29.66   &   3.5645 &   18.30 &  1.78  \\                        
J0209          &   02:09:44.61 &   +05:17:13.66   &   4.1846 &   18.44 &  1.58  \\                         
J0244          &   02:44:01.84 &   -01:34:03.78   &   4.044   &   18.18 &  1.62   \\                          
J0334          &   03:34:13.42 &   -16:12:05.36   &   4.380   &   18.63 &  1.66   \\                          
J0339          &   03:39:00.98 &   -01:33:17.70   &   3.204   &   19.17 &  1.61  \\                         
J0949          &   09:49:32.26 &   +03:35:31.78   &   4.1072 &   18.03 &  1.82  \\                          
J0958          &   09:58:52.19 &   +12:02:45.04   &   3.2746 &   17.47 &  2.95   \\                          
J1110          &   11:10:08.61 &   +02:44:58.07   &   4.1582 &   18.28 &  1.97   \\                       
J1111          &   11:11:13.79 &   -08:04:02.00   &   3.930   &   18.49 & 2.26 \\
J1209          &   12:09:17.93 &   +11:38:30.34   &   3.0836 &   17.45 &  1.88   \\                          
J1230          &   12:30:55.57 &   -11:39:09.30   &   3.557   &   19.84 &  1.75 \\                          
J1249          &   12:49:57.23 &   -01:59:28.80   &   3.6337 &   17.78 &  2.10 \\                          
J1332          &   13:32:54.51 &   +00:52:50.63   &   3.5071 &   18.35 &  1.62 \\                          
J1621          &   16:21:16.92 &   -00:42:50.86   &   3.7100 &   17.28 &  1.92   \\                          
J1939          &   19:39:57.25 &   -10:02:41.50   &   3.787   &   16.61 &  2.00  \\                          
J2003          &   20:03:24.14 &   -32:51:44.80   &   3.785   &   17.38 &  1.45  \\                          
J2053          &   20:53:44.72 &   -35:46:55.20   &   3.490   &   18.41 &  2.16   \\                         
J2215          &   22:15:27.29 &   -16:11:33.00   &   4.000   &   18.13 &  2.26   \\                           
J2303          &   23:03:01.45 &   -09:39:30.72   &   3.4774 &   17.68 &  1.75  \\                            
J2315          &   23:15:43.56 &   +14:56:06.41   &   3.3971 &   18.54 &  1.77  \\                          
J2334          &   23:34:46.40 &   -09:08:12.24   &   3.3261 &   18.03 &  1.66   \\                          
J2349          &   23:49:13.75 &   -37:12:59.25   &   4.240   &   19.15 &  1.56  \\                          
\hline
\end{tabular}
\end{adjustbox}
\footnotesize
\textsuperscript{a} Typical uncertainties range from 0.0001 to 0.013 \\
\textsuperscript{b} Typical photometric uncertainties range from 0.01 to 0.05 mag.\\
\end{table}

\begin{figure*}[t]
   \begin{center}
   \includegraphics[height=0.54\textheight,angle=0]{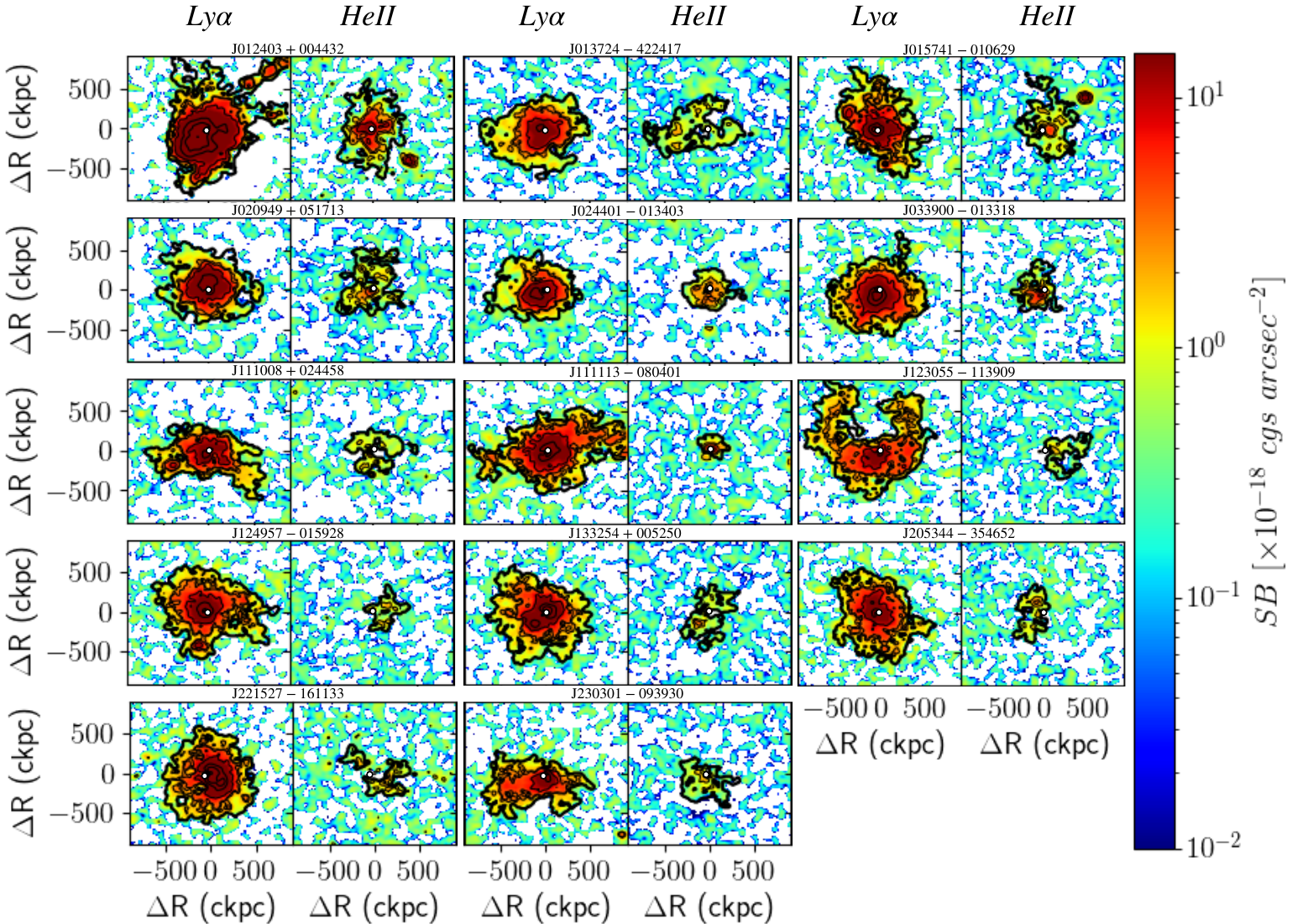} 
   \caption{Optimally extracted images in SB units ($10^{-18} \rm ~erg~s^{-1}~cm^{-2}~arcsec^2$) of the \lya\ and \heii\ emission lines tracing the CGM around 14 luminous quasars at the median redshift of $z \simeq 3.671$ (i.e., 500 ckpc corresponds to <$\approx 100~\rm pkpc$). The black thin contours mark the S/N levels 2.5, 3, 5 and 15, while the thick contours enclose the region in which each nebula is defined based on the z-direction collapsed 3D-mask obtained with the \texttt{CubEx} tool. The white dots mark the positions of the quasars. Each image shows a box 1.8$\times$1.8~$ \rm cMpc^2$.}
   \label{figure:SBmaps}
   \end{center}
\end{figure*}

Section~\ref{sec:method} describes the method to extract \lya\ and \heii\ nebulae in MUSE data. Figure~\ref{figure:SBmaps} reports these optimally extracted narrow-band images in SB units (i.e. $ \rm erg~s^{-1}~cm^{-2}~arcsec^{-2}$) of the \lya\ and \heii\ nebulae, in a box of size 1.8$\times$1.8~$\rm cMpc^2$. These images were smoothed, after the projection, with a Gaussian filter of radius 2 pixels.
The \texttt{Cube2Im} tool was also run to generate a S/N image of the extended emission line. In the maps in Figure~\ref{figure:SBmaps}, we display the thin black contours to show the S/N levels at 2.5, 3, 5, 10, and 15. The thick black contours outline the region defined by the collapsed 3D-mask, which corresponds to the nebula detection region identified by \texttt{CubEx}. Any spaxel not included in the projected 3D-mask was substituted with a median of the values extracted from the three layers at the wavelength centroid derived from the extended emission. The white dot denotes the position of the quasar, which represents the centroid of the PSF.

We performed two tests designed to verify the reliability of the \heii\ detections. These tests were based on the assumption that, once the sky and instrumental features are excluded, the overall flux attributed to random noise or PSF residuals detected with \texttt{CubEx}, and originating from local noise (e.g., PSF residuals), must be lower than the flux associated with a bona fide detection.

In the first test, we compared the flux of the \heii\ nebula emission to that of other extended objects detected using \texttt{CubEx} within the same datacube. We included only objects located at a distance less than 150 $\rm ckpc$ from the quasar position and within a range of $\pm 4000~\rm km~s^{-1}$ along the spectrum, relative to the centroid of the hypothetical \heii\ emission. 
In the second test, we performed a similar flux comparison between the \heii\ detections and the false \texttt{CubEx} detections (mostly PSF residuals) in other datacubes, within the same spectral window. However, we applied a selection process to choose the appropriate datacubes for this comparison. Specifically, for each datacube with a \heii\ detection, we selected datacubes that exhibit no emission lines within that spectral window. Additionally, we ensure that the standard deviation of the datacubes in the same spectral region is similar within a factor of 2. 

\begin{figure}[t]
   \begin{center}
   \includegraphics[height=0.25\textheight,angle=0]{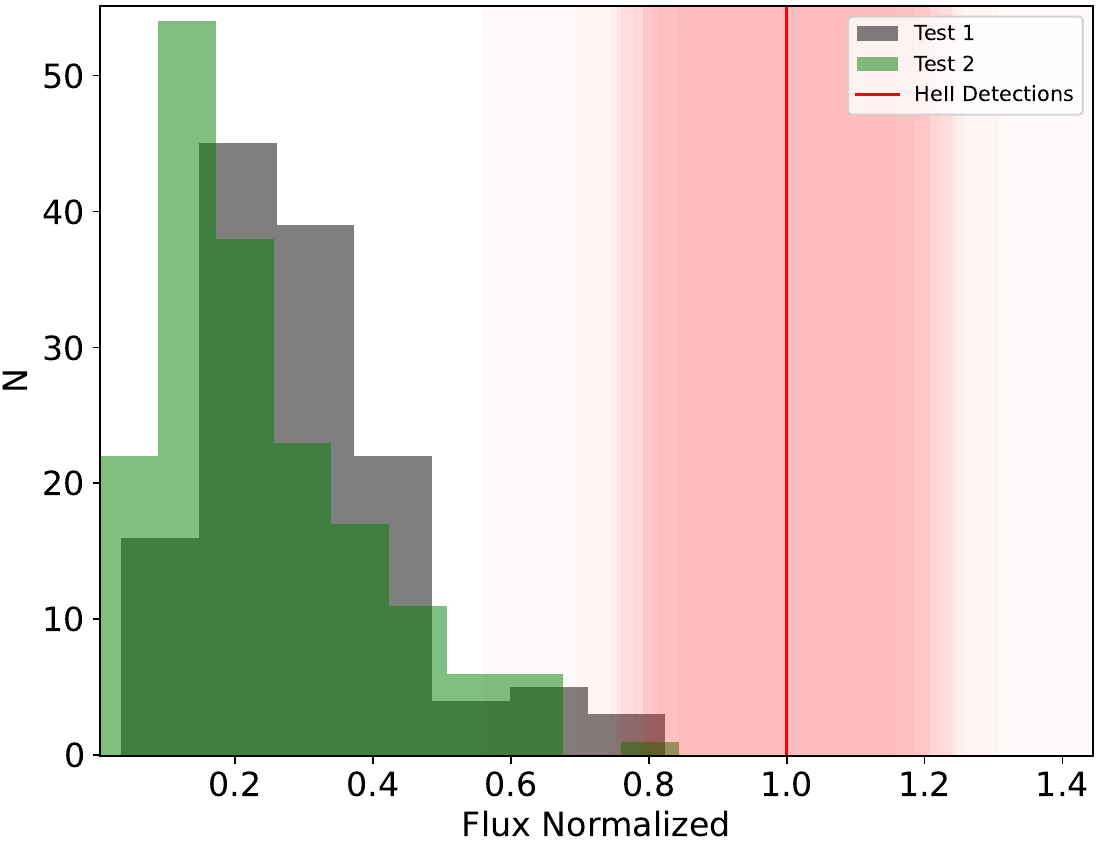}
   \caption{Histogram displaying the distribution of normalized fluxes for the \heii\ detections reported as a red line, with the corresponding errors indicated by transparent red shading. Additionally, the fluxes for false detections normalized to the fluxes of the detections, identified by the \texttt{CubEx} method, are included in the histograms. Two different tests are performed to examine the false detections. In the first test (black), the false detections are collected from the same datacube, while in the second test (green), the false detections are obtained from different datacubes within the same spectral window as the \heii\ detection (see text for further details). It is important to note that the false detections included in the histograms exclude those that contain sky or instrumental features.}
   \label{figapp:stat1}
   \end{center}
\end{figure}
\setlength{\textfloatsep}{-2pt}

The last step of both tests is to verify that the shape of the candidate extended \heii\ nebula is not similar to those of the other extended objects and to visually inspect the datacube to confirm the \heii\ detection.

Figure~\ref{figapp:stat1} displays a histogram illustrating the distribution of fluxes of the \heii\ nebulae, which are represented by the red line and are normalized to a value of 1 for each respective detection. The transparent red shading represents the overlapping uncertainties of the normalized flux measurements for the \heii\ nebulae. To perform a correct comparison, the fluxes of the false detections obtained from test 1 (green) and test 2 (black) are also divided by the flux of the respective \heii\ detection. 
This histogram provides compelling evidence for the statistical significance of the detected \heii\ nebulae in the present study.
The 95th percentile of the distribution of the false detections' fluxes is $\sim$0.58, indicating that its distance from the \heii\ detections' fluxes is more than 3$\sigma$.

\begin{figure*}[t]
   \begin{center}
   \includegraphics[height=0.38\textheight,angle=0]{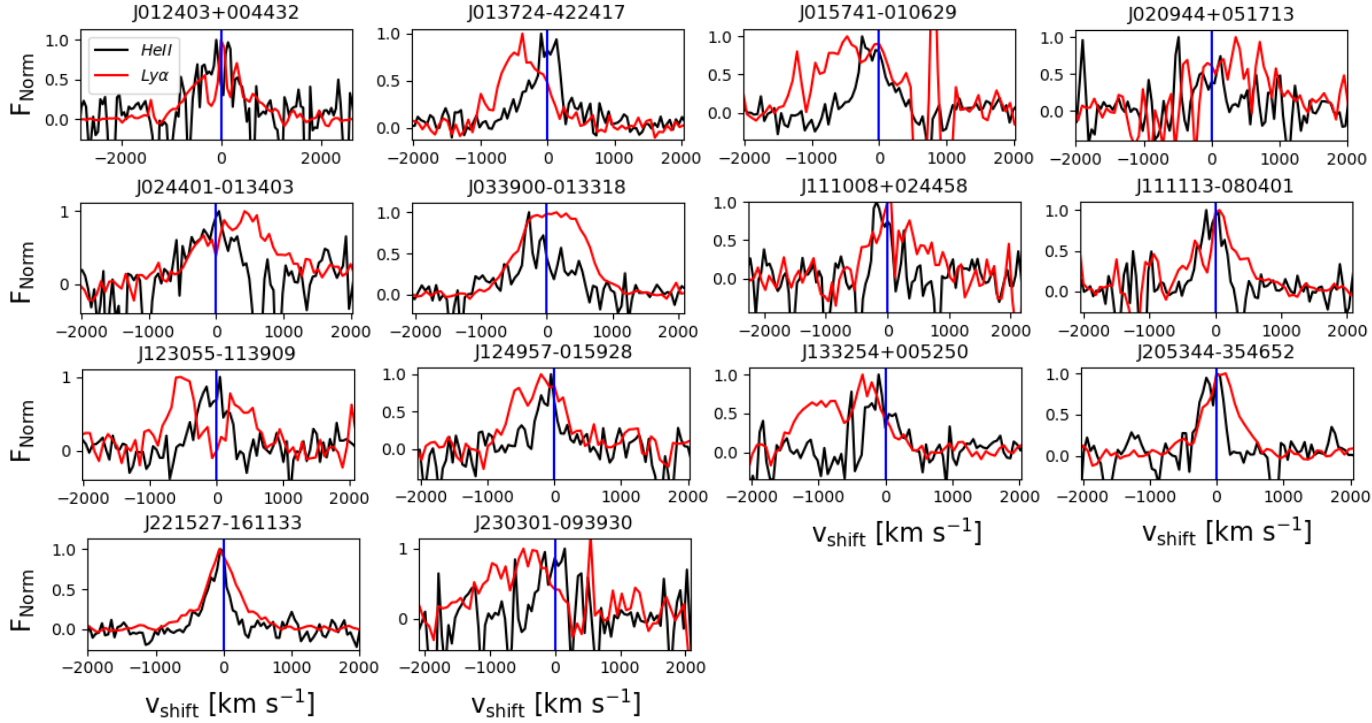}
   \caption{Plots of the \heii\ (black) and \lya\ (red) emission lines extracted from regions where both the \heii\ and \lya\ nebulae are detected. The data is displayed as a function of the velocity offset to the systemic redshift $\rm z_{HeII}$, with the S/N of the \heii\ emission line greater than 3. To avoid contamination from PSF residuals, we have excluded the central region within a circle of 3-pixel (0.6$\arcsec$) radius.}
   \label{fig:spec}
   \end{center}
\end{figure*}

Figure~\ref{fig:spec} presents spectra with a zoomed-in view of the \heii\ and \lya\ transitions. These spectra were extracted from regions where both nebulae are detected, and the S/N of the \heii\ emission line exceeds 3. To mitigate noise from PSF subtraction residuals, we have omitted the central portion within a circle of 2-pixel radius. The blue line denotes the wavelength corresponding to the peak of the total \heii\ line emission, serving as a reference for the nebula's redshift.

Table~\ref{tab:heii_lya_annuli_50_160} shows the dimming-corrected SB values and the \heii/\lya\ line ratios of the nebulae detected in KCWI and MUSE data, between 50 and 160 ckpc, presented in Figure~\ref{figure:SBratio}.

\begin{table}
\centering
\caption{Dimming-corrected SB and \heii/\lya\ ratios measured in the annulus $50 < R < 160$ ckpc. The SB is in units of $10^{-15}\,\mathrm{erg\,s^{-1}\,cm^{-2}\,arcsec^{-2}}$.}
\label{tab:heii_lya_annuli_50_160}
\tiny
\setlength{\tabcolsep}{1.6pt}
\renewcommand{\arraystretch}{0.92}
\resizebox{0.8\columnwidth}{!}{
\begin{tabular}{lccc}
\hline
Name &  $SB_{\lya}$ & $SB_{\heii}$ & $\heii/\lya$ \\
\hline
\multicolumn{4}{c}{Keck/KCWI}\\
\hline
CH-Q01  & $1.61 \pm 0.16$ & $0.02 \pm 0.08$ & $0.012 \pm 0.051$ \\
CH-Q02  & $1.93 \pm 0.26$ & $0.10 \pm 0.09$ & $0.054 \pm 0.048$ \\
J0010+06 & $1.65 \pm 0.09$ & $0.10 \pm 0.05$ & $0.060 \pm 0.030$ \\
J0338-06  & $1.56 \pm 0.24$ & $<0.13$ & $<0.084$ \\
Q0050+00 & $2.86 \pm 0.21$ & $0.41 \pm 0.10$ & $0.144 \pm 0.036$ \\
Q1227+28 & $7.95 \pm 0.26$ & $0.36 \pm 0.06$ & $0.045 \pm 0.007$ \\
Q1230+33 & $4.99 \pm 0.13$ & $0.44 \pm 0.28$ & $0.089 \pm 0.057$ \\
Q1411+00  & $22.67 \pm 0.32$ & $1.20 \pm 0.07$ & $0.053 \pm 0.003$ \\
Q1426+25  & $3.59 \pm 0.21$ & $0.32 \pm 0.08$ & $0.088 \pm 0.022$ \\
\hline
\multicolumn{4}{c}{VLT/MUSE} \\
\hline
J0124+00 & $20.13 \pm 0.23$ & $1.26 \pm 0.08$ & $0.062 \pm 0.004$ \\
J0137-42  & $4.29 \pm 0.06$ & $0.38 \pm 0.05$ & $0.090 \pm 0.011$ \\
J0157-01  & $3.04 \pm 0.05$ & $0.31 \pm 0.03$ & $0.102 \pm 0.011$ \\
J0209+05  & $7.66 \pm 0.08$ & $0.41 \pm 0.05$ & $0.054 \pm 0.006$ \\
J0244-01  & $4.37 \pm 0.07$ & $0.52 \pm 0.08$ & $0.119 \pm 0.019$ \\
J0339-01  & $3.33 \pm 0.04$ & $0.20 \pm 0.02$ & $0.061 \pm 0.006$ \\
J1110+02  & $4.27 \pm 0.12$ & $0.09 \pm 0.08$ & $0.020 \pm 0.019$ \\
J1111-08  & $6.34 \pm 0.09$ & $0.16 \pm 0.05$ & $0.025 \pm 0.008$ \\
J1230-11  & $1.73 \pm 0.08$ & $0.09 \pm 0.03$ & $0.052 \pm 0.017$ \\
J1249-01  & $3.77 \pm 0.07$ & $0.15 \pm 0.05$ & $0.040 \pm 0.013$ \\
J1332+00  & $4.30 \pm 0.05$ & $0.16 \pm 0.02$ & $0.036 \pm 0.005$ \\
J2053-35  & $2.92 \pm 0.07$ & $0.12 \pm 0.03$ & $0.041 \pm 0.009$ \\
J2215-16  & $8.63 \pm 0.10$ & $0.25 \pm 0.04$ & $0.029 \pm 0.005$ \\
J2303-09 & $1.52 \pm 0.04$ & $0.14 \pm 0.03$ & $0.089 \pm 0.018$ \\
\hline
\end{tabular}
}
\end{table}

\section{No detection of \heii\ nebulae}\label{app:noheii}

The depth of the observations in the MUSE sample allowed us to achieve a cosmological dimming-corrected SB limit of $\approx 8 \times 10^{-17}~\rm erg~s^{-1}~cm^{-2}~arcsec^{-2}$ at 1$\sigma$, within a 1$\rm ~arcsec^2$ aperture and per 1.25$~\AA$ observed-frame spectral layer.
This improved sensitivity has allowed us to detect \heii\ nebulae in $\sim 54 \%$ of the entire sample we analyzed, which is reported in Table~\ref{tabapp:magg}.
In contrast, other studies such as \citealt{Borisova16,Battaia19,Farina19}, which focused on \lya\ nebulae around luminous $z=3$-$4$ quasars, analyzed MUSE observations with exposure times of 1 hour or less. The SB limits in these observations, after correcting for cosmological dimming, were too high to detect potential \heii\ nebulae, except in individual extreme cases.
When comparing the SB limit to the expected radial profile of $\rm SB_{\heii}$ from previous samples, derived assuming the radial profile of $\rm SB_{\heii}/SB_{\lya}$ ratio obtained in our work, it becomes evident that, on average, these surveys are not sensitive enough to detect \heii\ nebulae. Given the measured ratio and the typical $\rm{SB}_{\lya} (1+z)^4$, a minimum SB limit that should be reached to have a high probability of detecting an \heii\ nebula is $\rm{SB}_{lim} \times (1+z)^4 \approx 2 \times 10^{-16}~\rm erg~s^{-1}~cm^{-2}~arcsec^{-2}$.\\

\begin{figure}[t]
   \begin{center}
   \includegraphics[height=0.28\textheight,angle=0]{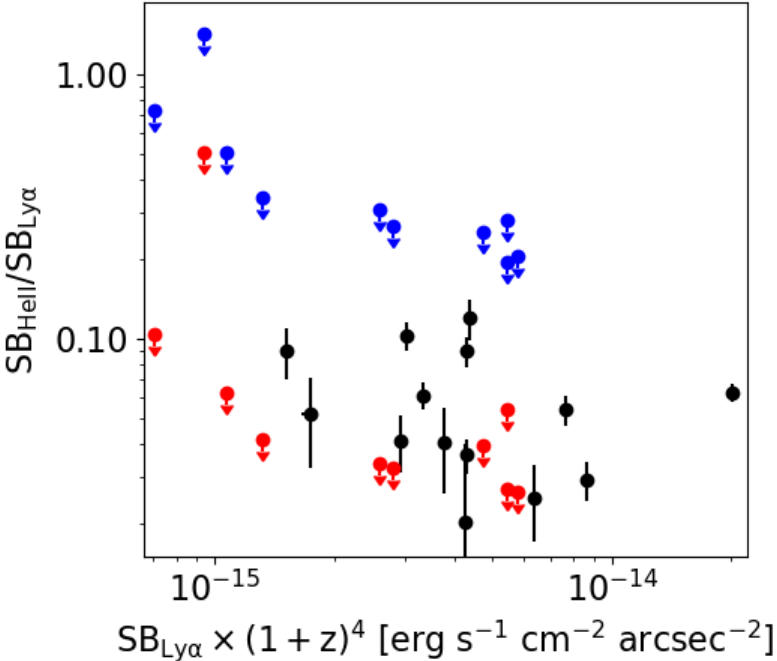}
   \caption{Cosmological dimming-corrected $\rm SB_{\lya}$ emission versus the $SB_{\heii}/SB_{\lya}$ ratios, and the upper limits estimated for the cases with no \heii\ detections. The upper limits are estimated under two different assumptions: the more conservative case in which the areas of the \heii\ and \lya\ nebulae are the same (blue); the area expected for the \heii\ nebulae is equivalent to the median of our \heii\ detection areas (red).}
   \label{figure:SBLyanoHeII}
   \end{center}
\end{figure}
\setlength{\textfloatsep}{-2pt}

Is there a physical or technical reason for the absence of \heii\ nebulae in almost half of the sample?
To address this question, we present in Figure~\ref{figure:SBLyanoHeII} the cosmological dimming-corrected $\rm SB_{\lya}$ emission plotted against the $\rm{SB}_{\heii}/\rm{SB}_{\lya}$ ratios, which are upper limit estimated when derived from the \lya\ nebulae where no \heii\ emission counterpart was detected. To estimate the upper limits for non-detections of \heii\ emission, we created a datacube cut with 40 layers, which is the maximum number of layers we observed in our \heii\ detections. The cut is centered on the wavelength corresponding to the expected \heii\ line emission based on the peak flux of the \lya\ line. By calculating the standard deviation of this datacube cut, we can determine the upper limits for non-detections of \heii\ emission. To do this, the standard deviation was multiplied by the square root of the aperture area, and divided by the square root of the average number of layers expected for the \heii\ extended emission, i.e., 5 layers.

The $F_{\heii}/F_{\lya}$ ratio upper limit is equal to $\rm SB_{\heii} ^{upper~limit} \times{\rm{Area}_{\heii}}/F_{\lya}$ where we considered, in a more conservative scenario, the area of the \heii\ nebula as large as the associated \lya\ nebula one (blue), otherwise it was considered as the median area of those of our \heii\ detections (red).
For comparison with these upper limits, we show the line ratios (black dots) estimated in Section~\ref{sec:ratios} as lower limits. This comparison is valid because both sets of ratios are estimated similarly, disregarding the effective area in which the \heii\ emission line is detected.
In the most conservative scenario (blue dots), the upper limits are above the line ratios derived for the \heii\ detections, as expected by construction.
In the scenario in which the area of the hypothetical \heii\ nebula is similar to the median area of all the \heii\ nebulae, we find that the upper limits are at the same level as the cases in which we have a \heii\ detection, some in the lower part of the distribution. This suggests that we are missing detections because the \heii\ nebulae could be fainter or more diffuse than our detections.

\end{document}